\documentclass[12pt,preprint]{aastex}
\DeclareGraphicsExtensions{.jpg,.pdf,.png,.eps,.ps}
\usepackage{graphics,graphicx,natbib}

\newcommand{\eps}[1]{\mbox{log~$\epsilon$(#1)}}

\def\cs22964{\mbox{CS~22964-161}}

\def\eg{\mbox{\it e.g.}}
\def\etal{\mbox{et al.}}

\def\logg{\mbox{log~{\it g}}}
\def\Msun{\mbox{$M_{\odot}$}}
\def\teff{\mbox{$T_{\rm eff}$}}
\def\vturb{\mbox{$v_{\rm t}$}}
\def\rpro{\mbox{$r$-process}}
\def\spro{\mbox{$s$-process}}
\def\hminus{H$^{-}$}
\def\ncap{\mbox{$n$-capture}}
\def\us{\char`\_}

\shorttitle{Neutron-Capture Elements in M15}
\shortauthors{Sobeck et al.}

\begin{document}

\title{
The Abundances of Neutron Capture Species in the Very Metal-Poor 
Globular Cluster M15: An Uniform Analysis of RGB and RHB Stars
}

\author{
Jennifer S. Sobeck\altaffilmark{1}, 
Robert P. Kraft\altaffilmark{2}, 
Christopher Sneden\altaffilmark{3}, 
George W. Preston\altaffilmark{4},
John J. Cowan\altaffilmark{5},
Graeme H. Smith\altaffilmark{2}, 
Ian B. Thompson\altaffilmark{4}, 
Stephen A. Shectman\altaffilmark{4}, 
Gregory S. Burley\altaffilmark{4}
}

\altaffiltext{1}{Department of Astronomy and Astrophysics, University
                 of Chicago, 5640 South Ellis Ave, Chicago, IL 60637; 
                 jsobeck@uchicago.edu}

\altaffiltext{2}{Department of Astronomy and Astrophysics, University
                 of California-Santa Cruz, Santa Cruz, CA 95064; 
                 kraft,graeme@ucolick.org}

\altaffiltext{3}{Department of Astronomy, University of Texas-Austin, 
                 1 University Station C1400, Austin, TX 78712; 
                 chris@verdi.as.utexas.edu}

\altaffiltext{4}{The Observatories of the Carnegie Institution of 
                 Washington, 813 Santa Barbara Street, Pasadena, CA 91101; 
                 gwp,ian,sshectman,burley@obs.carnegiescience.edu}

\altaffiltext{5}{Homer L. Dodge Department of Physics and Astronomy,
                 University of Oklahoma, Norman, OK 73019;
                 cowan@nhn.ou.edu}

\begin{abstract}

The globular cluster M15 is unique in its display of star-to-star variations in the neutron-capture elements.  Comprehensive abundance surveys 
have been previously conducted for handfuls of M15 red giant branch (RGB) and red horizontal branch (RHB) stars.  No attempt has been made to perform a single, 
self-consistent analysis of these stars, which exhibit a wide range in atmospheric parameters.  In the current effort, a new comparative abundance
derivation is presented for three RGB and six RHB members of the cluster.  The analysis employs an updated version of the line transfer code MOOG, which now 
appropriately treats coherent, isotropic scattering.  The apparent discrepancy in the previously reported values for the
metallicity of M15 RGB and RHB stars is addressed and a resolute disparity of $\Delta(RHB-RGB) \approx 0.1$ dex in the iron abundance was found.  The anti-correlative 
behavior of the light neutron capture elements (Sr, Y, Zr) is clearly demonstrated with both Ba and Eu, standard markers of the {\it s}- and \rpro, respectively.  
No conclusive detection of Pb was made in the RGB targets.  Consequently for the M15 cluster, this suggests that the main component of 
the \spro\ has made a negligible contribution to those elements normally dominated 
by this process in solar system material.  Additionally for the M15 sample, a large Eu abundance spread is confirmed, 
which is comparable to that of the halo field at the same metallicity.  These abundance results are considered in the discussion of the chemical 
inhomogeneity and nucleosynthetic history of M15. 
\end{abstract}

\section{INTRODUCTION\label{intro}}

Detections of multiple main sequences and giant branches in globular clusters (GC; \eg\ $\omega$Cen, NGC~2808, and NGC~1851; 
Bedin \etal\ 2004\citep{bed04}, Piotto \etal\ 2007\citep{pio07}, Han \etal\ 2009\citep{han09}) have challenged
the notion that these objects are {\it uniformly} mono-metallic stellar systems of unique age. 
In addition to the metallicity variations observed in certain clusters, the anomalous abundance behaviors of globulars 
include star-to-star scatter of light element [el/Fe] ratios (for C, N, O, Na, Mg, and Al) in both main sequence 
and giant stars (this is in contrast to the abundance trends of halo field stars; \eg\ Carretta \etal\ 2009a\citep{car09a} and Gratton, Sneden, \& 
Carretta 2004\citep{gra04})\footnote{We adopt the standard spectroscopic notation (Helfer, Wallerstein, \& Greenstein 1959\citep{hel59}) that 
for elements A and B, [A/B] $\equiv$ log$_{\rm 10}$(N$_{\rm A}$/N$_{\rm B}$)$_{\star}$ - log$_{\rm 10}$(N$_{\rm A}$/N$_{\rm B}$)$_{\odot}$.
We also employ the definition \eps{A} $\equiv$ log$_{\rm 10}$(N$_{\rm A}$/N$_{\rm H}$) + 12.0.}.  These departures in 
the relative abundances (found in stars of different evolutionary stages) imply that there are
multiple stellar generations present within the globular cluster and that an initial generation may have
contributed to the intracluster medium.  It is possible that three sources are responsible for the 
aggregate chemical makeup of a globular cluster: a {\it primordial} source that generates the initial composition of
the protocluster cloud, a {\it pollution} source that deposits material into the ICM from highly-evolved 
asymptotic branch stars, and a {\it mixing} source that is independent of the other two and the
result of stellar evolution processes.  Further discussion of these scenarios may be found in, \eg\, 
Bekki et al. (2007)\citep{bek07} and Carretta et al. (2009b)\citep{car09b}.   

On the other hand in the vast majority of globular clusters, minimal scatter in the element abundance ratios with Z$>$20 has been observed.
The abundances for the neutron(n-) capture elements europium and barium have been measured in several GC's, and only in a few, exceptional cases have 
significantly large intracluster differences in these values been seen (\eg\ M22; Marino et al. 2009\citep{mar09}).  The predominant mechanism
of Eu manufacture is the rapid n-capture process (\rpro) whereas the primary nucleosynthetic channel for Ba is
slow n-capture (\spro; additional information pertaining to these production mechanisms may be found in 
\eg\ Sneden, Cowan, \& Gallino 2008\citep{sne08}).  Consequently, the abundance ratio of [Eu/Ba] is used to 
demonstrate the relative prevalence of the $r$- or \spro\ in individual stars.
In globular clusters with a metallicity of [Fe/H]$\lesssim -1$, a general enhancement of [Eu/Ba]~$\sim$~+0.4 to +0.6 dex is detected, which
indicates that n-capture element production has been dominated by the \rpro\ (Gratton, Sneden, \& Carretta 2004\citep{gra04} and references therein).  This in 
turn is suggestive of explosive nucleosynthetic input from very massive stars.         

The very metal-deficient globular cluster M15 (NGC 7078; [Fe/H]$\sim -2.3$) has been subject
to several abundance investigations including the recent study by Carretta et al. (2009a\citep{car09a}).
They employed both medium-resolution and high-resolution spectra of over 80 red giant stars to precisely determine
the metallicity of this cluster: $<$[Fe/H]$> = -2.314 \pm 0.007$.  Additionally, they detected variations in the light 
element abundances (Na and O) for stars along the entirety of the Red Giant Branch (RGB).  Prior studies
have also observed large scatter in the relative Ba and Eu (intracluster) abundances.  With the spectra of 17 RGB stars,
Sneden et al. (1997\citep{sne97}) found a factor of three spread in both ratios: $<$[Ba/Fe]$> = 0.07$; $\sigma= 0.18$ and 
$<$[Eu/Fe]$> = 0.49$; $\sigma= 0.20$.\footnote{The anomalously nitrogen-enriched star K969 is omitted; see Appendix A of Sneden \etal\ 1997\citep{sne97}}
  They were able to exclude measurement error as the source for the scatter and determined 
that the variations were correlated: $<$[Eu/Ba]$>= 0.41$; $\sigma= 0.11$.  In a follow-up study of 31 M15 giants by Sneden, 
Pilachowski \& Kraft (2000b\citep{sne00b}), the scatter in the relative Ba abundance was confirmed: $<$[Ba/Fe]$>= 0.12$; $\sigma = 0.21$ 
(limitations in the spectral coverage did not permit a corresponding analysis of Eu).  

The majority of M15 high-resolution abundance analyses have employed yellow-red visible spectra to maximize signal-to-noise 
(stellar flux levels are relatively high for RGB targets in this region).  In order to precisely derive the neutron capture abundance distribution in M15, 
Sneden et al. (2000a\citep{sne00a}) re-observed three tip giants in the blue visible wavelength regime 
(which contains numerous n-capture spectral transitions).  The abundance determinations of 8 n-capture 
species (Ba, La, Ce, Nd, Sm, Eu, Gd and Dy) were performed and large star-to-star scatter in the all of the [El/Fe] ratios was measured.  They 
also found that the three stars exhibited a scaled solar system \rpro\ abundance pattern.  Additional verification of these abundance 
results was done by Otsuki \etal\ (2006\citep{ots06}) in an analysis of six M15 RGB stars (the two studies had one star in common, K462).  Consistent with
Sneden et al. (2000a\citep{sne00a}), they detected significant variation in the [Eu/Fe], [La/Fe] and [Ba/Fe] ratios.  Furthermore, Otsuki et al. found 
that the ratios of [(Y,Zr)/Eu] show distinct anticorrelations with the Eu abundance.  Finally employing an alternate stellar sample, Preston \etal\ (2006\citep{pre06}) 
examined six red horizontal branch (RHB) stars of M15.\footnote{The papers from Sneden et al. (1997), Sneden, 
Pilachowski, \& Kraft (2000b), Sneden et al. (2000a), and Preston et al. (2006) are from collaborators affiliated with institutions in both California and Texas.  
Hereafter, these papers and other associated publications will be referred to as CTG.}  For the elements Sr, Y, Zr, Ba and Eu
a large (star-to-star) spread in the abundances was measured.  In essence, all of these investigations have observed considerable chemical inhomogeneity in
the n-capture elements of the globular cluster M15.

Two issues are brought to light by the M15 abundance data: the timescale and efficiency of mixing in the protocluster environment; and, the 
nucleosynthetic mechanism(s) responsible for n-capture element manufacture.  In this globular cluster, large abundance variations are seen in the two 
stellar evolutionary classes as well as in both the light and heavy neutron capture species.  There is a definitive enhancement of \rpro\ elements 
found in some stars of M15 (\eg\, K462), yet not exhibited in others (\eg\, B584).  Taking into consideration the entirety of the M15 n-capture results, these data 
hint at the existence of a nucleosynthetic mechanism different from the classical {\it r-} and {\it s-}processes.  
Evidence of such a scenario (with multiple production pathways) may be also found in halo field stars of similar metallicity such 
as CS 22892-052 (Sneden \etal\ 2003\citep{sne03}) and HD 122563 (Honda \etal\ 2006\citep{hon06}), which have displayed similar abundance variations.  
Indeed, several models have advanced the notion of more than one \rpro\ formation scenario (\eg\ Wasserburg \& Qian 2000, 2002\citep{was00,was02}, 
Thielemann \etal\ 2001\citep{thi01}, and Kratz \etal\ 2007\citep{kra07}).

To further understand the implications of the M15 results, the spectra from the three RGB stars of Sneden et al. (2000a)\citep{sne00a}
and the six RHB stars of Preston \etal\ (2006)\citep{pre06} are re-analyzed.  A single consistent methodology for the analysis is employed and an 
expansive set of recently-determined oscillator strengths is utilized (\eg\ Lawler \etal\ 2009\citep{law09}, Sneden \etal\
2009\citep{sne09}, and references therein).  As the pre- and post-He core flash giants are examined, the relative invariance of abundance 
distributions will be ascertained for $r-$ and \spro\ species.  In consideration of the M15 investigations cited above, a few data anomalies have come to light.  
The two main issues to be resolved include: large discrepancies in the log~$\epsilon(El)$ values between the studies of Sneden et al. (2000a)\citep{sne00a} and 
Otsuki et al. (2006)\citep{ots06}; and, the significant disparity in the derived metallicity for the M15 cluster between Preston et al. 2006 ($<$[Fe/H]$>_{RHB} = -2.63$) 
and the canonically accepted value of $<$[Fe/H]$>= -2.3$ (\eg\ Carretta et al. 2009a\citep{car09a}, Sneden et al. 2000a\citep{sne00a}).  It is suggested that these 
differences are mostly due to selection of atomic data, model atmosphere, and treatment of scattering. 

\section{OBSERVATIONAL DATA\label{obs}}

For the three RGB stars of the M15 cluster, the re-analysis of two sets of high resolution spectra was performed: the first from 
Sneden \etal\ (1997\citep{sne97}) with approximate wavelength coverage region of 5400\AA~$\lesssim$ $\lambda$ $\lesssim$ 6800\AA\ and the second from 
Sneden \etal\ (2000a\citep{sne00a}) with a wavelength domain of 3600\AA~$\lesssim$ $\lambda$ $\lesssim$ 5200\AA.  All spectral observations were
acquired with the High Resolution Echelle Spectrometer (HIRES; Vogt et al. 1994\citep{vog94}) at the Keck I 10.0-m telescope (with 
a spectral resolving power of R $\equiv$ $\lambda$/$\Delta\lambda$ $\simeq$ 45000).  The signal-to-noise (S/N) range of the data varied from 
30 $\lesssim$ S/N $\lesssim$ 150 for the shorter wavelength spectra to 100 $\lesssim$ S/N $\lesssim$ 150 for the longer wavelength
spectra (the S/N value generally increased with wavelength).  The three giants, K341, K462, and K583\footnote{The Kustner
(1921)\citep{kus21} designations are employed throughout the text.}, were selected from the larger stellar sample of Sneden \etal\ (1997)\citep{sne97} 
due to relative brightness, rough equivalence of model atmospheric parameters, and extreme spread in associated Ba and Eu abundances.

Re-examination of the high resolution spectra of six RHB stars from the study of Preston \etal\ (2006)\citep{pre06} was also done.  The observations
were taken at the Magellan Clay 6.5-m telescope of the Las Campanas Observatory with the Magellan Inamori Kyocera Echelle (MIKE) 
spectrograph (Bernstein \etal\ 2003)\citep{ber03}.  The data had a resolution of R$\simeq$ 40000 and the S/N values ranged from 
$S/N\sim$25 at 3600~\AA\ to $S/N\sim$120 at 7200~\AA\ (note that almost complete spectral coverage was 
obtained in the region 3600\AA~$\lesssim$ $\lambda$ $\lesssim$ 7200\AA).  The six RHB targets 
were chosen from the photometric catalog of Buonanno \etal\ (1983)\citep{buo83} and accordingly signified as B009, B028, B224,
B262, B412 and B584.  It should be pointed out that these stars have significantly lower temperatures than other HB members (and thus,
match up favorably with the RGB).  

\begin{figure}
\centering%
\includegraphics[width=5.9in]{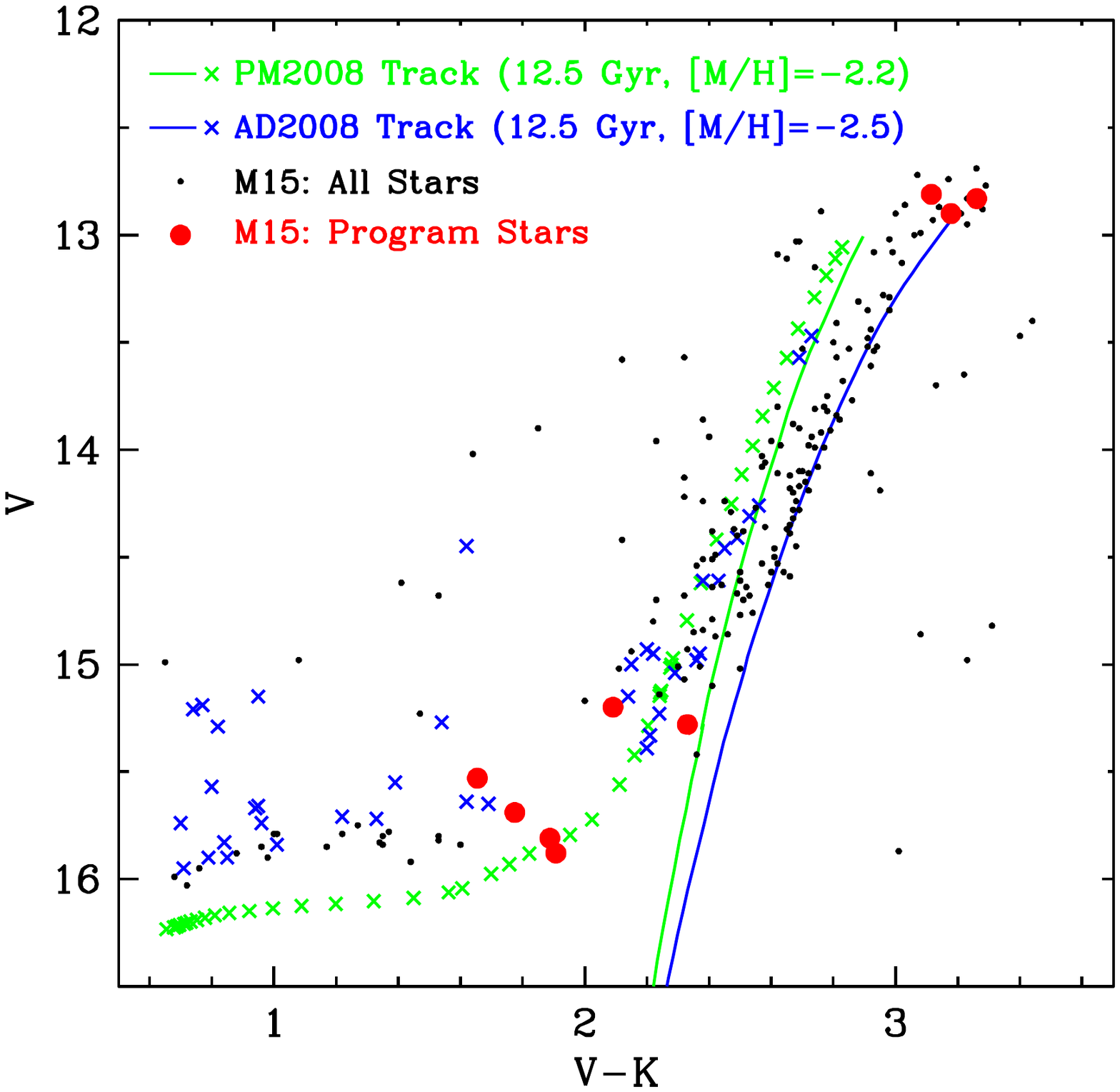}
\caption{A color-magnitude diagram for the globular cluster M15.  For cluster members (indicated by small black dots), $V$ magnitudes are taken Buonanno \etal\ (1983) while
the $K$ magnitudes are obtained from the 2MASS database (Skrutskie \etal\ 2006).  Note that stars with both $B$ and $V$ magnitudes from Buonanno \etal\ 
are shown in the plot.  The program stars are signified by large, red circles.  Also displayed are the isochrone data from 
both Marigo et al. (2008; labeled PM2008; shown in green) and Dotter et al. (2008; labeled AD2008; shown in blue).}  
\label{vminusk}
\end{figure}

Figure~\ref{vminusk} features the color-magnitude diagram (CMD) for the M15 globular cluster with a plot of the $V$ versus $(V-K)$ magnitudes.   The $V$ 
magnitudes for the RGB stars are taken from the prelimiary results of Cudworth (2011) and verified against the data from Cudworth (1976\citep{cud76}).  
Alternatively, the RHB $V$ magnitude values are obtained from Buonanno et al. (1983\citep{buo83}).  The $K$ magnitudes for all M15 targets are taken from the 
Two Micron All Sky Survey (2MASS; Skrutskie \etal\ 2006\citep{skr06}).  Cluster members 
with both $B$ and $V$ measurements from Buonanno \etal\ are displayed in the plot (denoted by the black dots) and the stars of the current study 
are indicated by large, red circles.  Note that the identifications of RGB and RHB members are based upon stellar atmospheric parameters as well as
the findings from Sneden \etal\ (1997\citep{sne97}) and Preston \etal\ (2006\citep{pre06}; please consult those references for additional details).  
Also in Figure~\ref{vminusk}, two isochrone determinations are overlayed upon the photometric data: Marigo \etal\ (2008\citep{mar08}; 
with the age parameter set to 12.5 Gyrs and a metallicity of [M/H]=-2.2; shown in green) and Dotter \etal\ (2008\citep{dot08}; with the age 
parameter set to 12.5 Gyrs and a metallicity of [M/H]=-2.5; shown in blue).  These are the best-fit isochrones to the general characteristics ascribed to M15 and 
no preference is given to either source.  

Additional observational details of the aforementioned data samples may be found in the original Sneden \etal\ (1997,2000a)\citep{sne97,sne00a} 
and Preston \etal\ (2006)\citep{pre06} publications.  These papers also contain descriptions of the data reduction procedures, 
in which standard IRAF\footnote{\footnotesize IRAF is distributed by NOAO, which is operated by AURA, under cooperative agreement with the
NSF.} tasks were used for extraction of multi-order wavelength-calibrated spectra from the raw data frames, and specialized software 
(SPECTRE; Fitzpatrick \& Sneden 1987\citep{fit87}) was employed for continuum normalization and cosmic ray elimination.

\begin{figure}
\centering%
\includegraphics[width=5.9in]{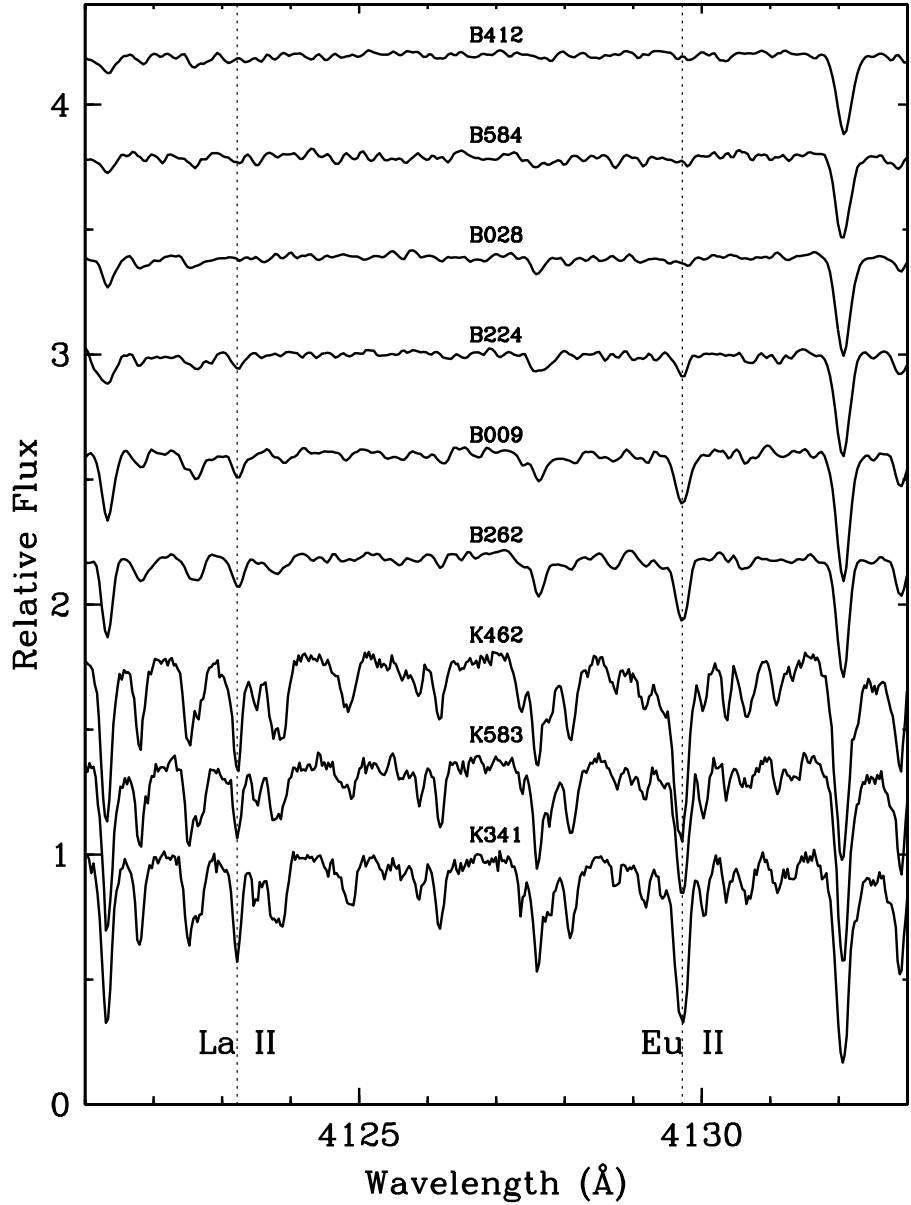}
\caption{A small wavelength region shown for all program stars.  For the purposes of display, the relative fluxes of target stars (other than K341)
have been shifted by additive constants.  Vertical dotted lines denote the spectral features of \ion{La}{2} at a wavelength of 4123.22~\AA\ 
and of \ion{Eu}{2} at 4129.72~\AA.  Note that these transitions appear to be more pronounced in the spectra of the lower temperature RGB targets.}
\label{spec1}
\end{figure}

Figure~\ref{spec1} features a comparison of the spectra of all M15 targets.  Displayed in this plot is a small wavelength interval $4121-4133$~\AA, 
which highlights the important n-capture transitions \ion{La}{2} at 4123.22~\AA\ and \ion{Eu}{2} at 4129.72~\AA.  The spectra are arranged
in decreasing \teff\ from the top to the bottom of the figure.  As shown, the combined effects of \teff\ and \logg\ influence the apparent line strength, and 
accordingly, transitions which are saturated in the RGB spectra completely disappear in the warmer RHB spectra.

\section{METHODOLOGY AND MODEL DETERMINATION}\label{modparam}

Several measures were implemented in order to improve and extend the efforts of Sneden et al. (1997, 2000a)\citep{sne97,sne00a} and Preston et al. 
(2006\citep{pre06}).  First, the modification of the line analysis program MOOG was performed to accurately ascertain the relative contributions
to the continuum opacity (especially necessary for the bluer wavelength regions and the cool, metal-poor RGB targets).  
Second, the employment of an alternative grid of models was done to obtain an internally consistent set of stellar atmospheric 
parameters for the total M15 sample.  Third, the utilization of the most up-to-date experimentally and semi-empirically-derived 
transition probability data was done to determine the abundances from multiple species. 

\subsection{Atomic Data}\label{atdata}

Special effort was made to employ the most recent laboratory measurements of oscillator strengths.  When applicable, the inclusion of 
hyperfine and isotopic structure was done for the derivation of abundances.  Tables 2 and 3 list the various literature sources for 
the transition probability data.  Some species deserve special comment.  The Fe transition probability values are taken from 
the critical compilation of Fuhr \& Wiese (2006\citep{fuh06}; note for neutral Fe, the authors heavily weigh the laboratory data from O'Brian \etal\ 
1991\citep{obr91}).  No {\it up-to-date} laboratory work has been done for Sr, and so, the adopted gf-values are from the 
semi-empirical study by Brage et al. (1998\citep{bra98}; these values are in good agreement with those derived empirically 
by Gratton \& Sneden 1994\citep{gra94}).  Similarly, the most recent laboratory effort for Y was by Hannaford \etal\ (1982\citep{han82}).  Yet 
these transition probabilities appear to be robust, yielding small line-to-line scatter.

A particular emphasis of the current work is the n-capture element abundances, for which a wealth of new transition probability data
have become recently available.  Correspondingly, the extensive sets of rare earth $gf$-values from the Wisconsin Atomic Physics Group were adopted 
(Sneden \etal\ 2009\citep{sne09}, Lawler et al. 2009\citep{law09}, and references therein).  These data when applied to the solar spectrum yield photospheric 
abundances that are in excellent agreement with meteoritic abundances.  For neutron capture elements not studied by 
the Wisconsin group (which include Ba, Pr, Yb, Os, Ir, and Th), alternate literature references 
were employed (and these are accordingly given in the two aforementioned tables). 

\subsection{Consideration of Isotropic, Coherent Scattering}

In the original version of the line transfer code MOOG (Sneden 1973\citep{sne73}), local thermodynamic equilibrium (LTE) was assumed and hence, 
scattering was treated as pure absorption.  Accordingly, the source function, $S_{\lambda}$, was set equal to the Planck function, $B_{\lambda}(T)$, which is an adequate 
assumption for metal-rich stars in all wavelength regions.  However for the extremely metal-deficient, cool M15 giants, 
the dominant source of opacity switches from  H$^{-}$$_{BF}$ to Rayleigh scattering in the blue 
visible and ultraviolet wavelength domain ($\lambda \lesssim 4500$\AA). It was then necessary to modify the MOOG program as the LTE approximation was no 
longer sufficient (this has also been remarked upon by other abundance surveys, \eg\ Johnson 2002\citep{joh02} and Cayrel \etal\ 2004\citep{cay04}).

The classical assumptions of one-dimensionality and plane-parallel geometry continue to be employed in the code.  Now with the
inclusion of isotropic, coherent scattering, the framework for solution of the radiative transfer equation (RTE) shifts from
an initial value to a boundary value problem.  The source function then assumes the form\footnote
{To re-state, the equation terms are defined as follows: $S$ is the source function, $\epsilon$ is the thermal
coupling parameter, $J$ is the mean intensity, and $B$ is the Planck function.} of $S = (1-\epsilon)J + \epsilon B$ 
and the description of line transfer becomes an integro-differential equation.  The chosen methodology for the
solution of the RTE (and the determination of mean intensity) is the approach of {\it short characteristics} that incorporates aspects of an 
accelerated convergence scheme.  In essence, the short characteristics
technique employs a tensor product grid in which the interpolation of intensity values occurs at selected grid points.  
The prescription generally followed was that from Koesterke et al. (2002\citep{koe02} and references therein).  The Appendix provides
more detail with regard to the MOOG program alterations.

Prior to these modifications, for a low temperature and low metallicity star (\eg\ a RGB target), the ultraviolet and blue visible spectral transitions 
reported aberrantly high abundances in comparison to those abundances found from redder lines.  With the implementation of the
revised code, better line-to-line agreement is found and accordingly, the majority of the abundance trend with wavelength is eliminated for 
these types of stars.  Note for the RHB targets, minimal changes are seen in abundances with the employment of the modified MOOG program
(as the dominant source of opacity for these relatively warm stars is always H$^{-}$$_{BF}$ over the spectral region of interest).

\subsection{Atmospheric Parameter Determination}

To obtain preliminary estimates of \teff\ and \logg\ for the M15 stars, photometric data from the aforementioned sources (Cudworth 2011;
Buonanno et al. 1983\citep{buo83}; 2MASS) were employed as well as those data from Yanny et al. 1994\citep{yan94}.  To transform the color, 
the color-\teff\ relations of Alonso \etal\ (1999)\citep{alo99} were used in conjunction with the distance modulus ($(m-M)_0$ = 15.25) and 
reddening ($E(B-V)$~=~0.10) determinations from Kraft \& Ivans (2003\citep{kra03}).  Note that an additional intrinsic uncertainty of about 
0.1~dex in \logg\ remains among luminous RGB stars owing to stochastic mass loss of order 0.1~dex.  Consequently, initial masses of 0.8~\Msun\ and 0.6\Msun\ were assumed 
for RGB and RHB stars respectively.  The photometric V and (V-K) values as well as the photometrically- and spectroscopically-derived stellar atmospheric parameters 
are collected in Table~\ref{m15models}.

With the use of the spectroscopic data analysis program SPECTRE (Fitzpatrick \& Sneden 1987\citep{fit87}), the equivalents widths (EW) of
transitions from the elements Ti I/II, Cr I/II, and Fe I/II were measured in the wavelength range 3800-6850 \AA.  The preliminary \teff\ 
values were adjusted to achieve zero slope in plots of Fe abundance (log~$(\epsilon_{Fe I}$)) as a function of excitation potential ($\chi$)
and wavelength ($\lambda$).  The initial values of \logg\ were tuned to minimize the disagreement between the neutral and ionized species 
abundances of Ti, Cr, and Fe (particular attention was paid to the Fe data).  Lastly, the microturbulent velocities \vturb\ were set as 
to reduce any dependence on abundance as a function of EW.  Final values of \teff, \logg, \vturb, and metallicity [Fe/H] are 
listed in Table~1, as well as those values previously derived by Sneden \etal\ (2000a\citep{sne00a}) for the RGB stars and 
Preston \etal\ (2006)\citep{pre06} for the RHB stars. 

\subsection{Selection of Model Type}

To conduct a standard abundance analysis under the fundamental assumptions of one-dimensionality and local thermodynamic equilibrium (LTE), two 
grids of model atmospheres are generally employed: Kurucz-Castelli (Castelli \& Kurucz 2003\citep{cas03}; Kurucz 2005\citep{kur05}) and 
MARCS (Gustafsson \etal\ 2008\citep{gus08}).\footnote{Kurucz models are available through the website: http://kurucz.harvard.edu/
and MARCS models can be downloaded via the website: http://marcs.astro.uu.se/} The model selection criteria were as follows: the reconciliation 
of the metallicity discrepancy between the RGB and RHB stars of M15, the derivation of (spectroscopic-based) atmospheric parameters in 
reasonable agreement with those found via photometry, and the attainment of ionization balance between the Fe I and Fe II transitions.  
For the RHB targets, interpolated models from the Kurucz-Castelli and MARCS grids were comparable and yielded extremely similar abundance results.  
However, there are a few notable differences between the two model types for the RGB stars with regard to the P$_{gas}$ 
and P$_{electron}$ content pressures.  Though beyond the scope of the current effort, it would be of considerable interest to examine in detail the exact 
departures between the Kurucz-Castelli and MARCS grids.  To best achieve the aforementioned goals for the M15 data set, MARCS models were accordingly chosen. 

\subsection {Persistent Metallicity Disagreement between RGB and RHB stars}

For the RHB stars, the presently-derived metallicties differ slightly from those of 
Preston et al. (2006): $<$[Fe$_{I}$/H]$>$ = -2.69 (a change of $\Delta = -0.03$) and $<$[Fe$_{II}$/H]$>$ = -2.64 (a change of $\Delta = -0.04$).  
However for the RGB stars, the [Fe/H] results of the current study {\it do} vary 
significantly from those of Sneden et al. (2000): $<$[Fe$_{I}$/H]$>$ = -2.56 
(a downwards revision of $\Delta = -0.26$) and $<$[Fe$_{II}$/H]$>$ = -2.53 (a downwards revision of $\Delta = -0.28$).  The 
remaining metallicity discrepancy between the RGB and RHB stars is as follows: $\Delta(RGB-RHB)_{Fe I}$ = 0.13 and 
$\Delta(RGB-RHB)_{Fe II}$ = 0.11.  Even with the employment of MARCS models and the incorporation of 
Rayleigh scattering (not done in previous efforts), the offset persists.  Repeated exercises
with variations in the \teff, \logg, and \vturb\ values showed that this metallicity disagreement in all likelihood cannot be attributed
to differences in these atmospheric parameters.  As a further check, the derivation of [Fe$_{I,II}$/H] values was performed
with a list of transitions satisfactorily measurable in both RGB and RHB spectra.  No reduction in the metallicty disagreement was seen as 
the offsets were found to be: $\Delta(RGB-RHB)_{Fe I}$ = 0.11 (with 45 candidate Fe$_{I}$ lines) and 
$\Delta(RGB-RHB)_{Fe II}$ = 0.14 (with 3 candidate Fe II$_{II}$ lines).  

The data from the M15 RGB and RHB stars originate from different telescope/instrument set-ups. Additionally, somewhat different data 
reduction procedures were employed for the two samples.  Possible contributors to the iron abundance offset could be the lack of consideration of spherical symmetry 
in the line transfer computations and the generation of sufficiently representative stellar atmospheric models for these highly evolved stars 
(which exist at the very tip of the giant branch and/or have undergone He-core flash).  Indeed, it is difficult to posit a {\it single, clear-cut} explanation for 
the disparity in the RGB and RHB [Fe/H] values. In an analysis of the globular cluster M92, King \etal\ (1998\citep{kin98}) derived an average abundance 
ratio of $<$[Fe/H]$>= $ -2.52 for six subgiant stars, a factor of two lower than the $<$[Fe/H]$>$ value 
measured in the red giant stars.  Similarly, Korn \etal\ (2007)\citep{kor07} surveyed turn off (TO) and RGB stars of NGC~6397 and found 
a metallicity offset of about 0.15~dex (with the TO stars reporting consistently lower values of [Fe/H]). They 
argued that the TO stars were afflicted by gravitational settling and other mixing processes and as a result, the Fe abundances of 
giant stars were likely to be nearer to the {\it true} value. While the TO stars do have \teff\ values close to that of the M15 RHB stars, 
they have surface gravities and lifetimes that are considerably larger.  Accordingly, it is not clear if the offset in M15 has a 
physical explanation similar to that proposed in the case of NGC 6397.

\section{ABUNDANCE RESULTS}

For the extraction of abundances, the two filters of line strength and contaminant presence were used to assemble an effective line list.  Abundance derivations
for the majority of elements employed the technique of synthetic spectrum line profile fitting (accomplished with the updated MOOG code as described in \S3.2). 
For a small group of elements (those whose associated spectral features lack both hyperfine and isotopic structure), the simplified approach of EW measurement
was used (completed with both the MOOG code and the SPECTRE program; Fitzpatrick \& Sneden 1987\citep{fit87}).  Presented in Tables~\ref{m15rgbdata} and~\ref{m15rhbdata} 
are the log~$\epsilon$ abundance values for the individual transitions detected in the M15 RGB and RHB stars respectively.  These tables also list the 
relevant line parameters as well as the associated literature references for the $gf$-values employed.  

In addition to the line-to-line scatter, errors in the abundance results may arise due to uncertainties in the model atmospheric
parameters.  To quantify these errors in the M15 data set, a RGB target, K462, is first selected.  If alterations of $\Delta T_{eff}= \pm 100$ K 
are applied, then the abundances of neutral species change by approximately $\Delta$[X$_{I}$/H] $\simeq \pm 0.15$ whereas the 
abundances of singly-ionized species change by about $\Delta$[X$_{II}$/H] $\simeq \pm 0.04$.  Variations of $\Delta$\logg\ $= \pm 0.20$ yield 
$\Delta$[X$_{I}$/H] $\simeq \pm 0.04$ in neutral species abundances and $\Delta$[X$_{II}$/H] $\simeq \pm 0.05$ in
singly-ionized species abundances.  Changes in the microturbulent velocity on the order of $\Delta$\vturb\ $= \pm 0.20$ 
result in abundance variations of $\Delta$[X$_{I}$/H] $\simeq \pm 0.08$ and $\Delta$[X$_{II}$/H] $\simeq \pm 0.04$.  The exact same procedure is then 
repeated for the RHB star, B009.  Modifications of the temperature by $\Delta T_{eff}= \pm 100$ K lead to abundance 
changes of $\Delta$[X$_{I}$/H] $\simeq \pm 0.09$ in neutral species and $\Delta$[X$_{II}$/H] $\simeq \pm 0.02$ in singly-ionized species.  Alterations of 
the surface gravity by $\Delta$\logg\ $= \pm 0.20$ engender variations of  $\Delta$[X$_{I}$/H] $\simeq \pm 0.01$ and 
$\Delta$[X$_{II}$/H] $\simeq \pm 0.07$.  Finally, variations of $\Delta$\vturb\ $= \pm 0.20$ produce abundance changes of 
$\Delta$[X$_{I}$/H] $\simeq \pm 0.08$ and $\Delta$[X$_{II}$/H] $\simeq \pm 0.04$.

\begin{figure}
\centering%
\includegraphics[width=5.9in]{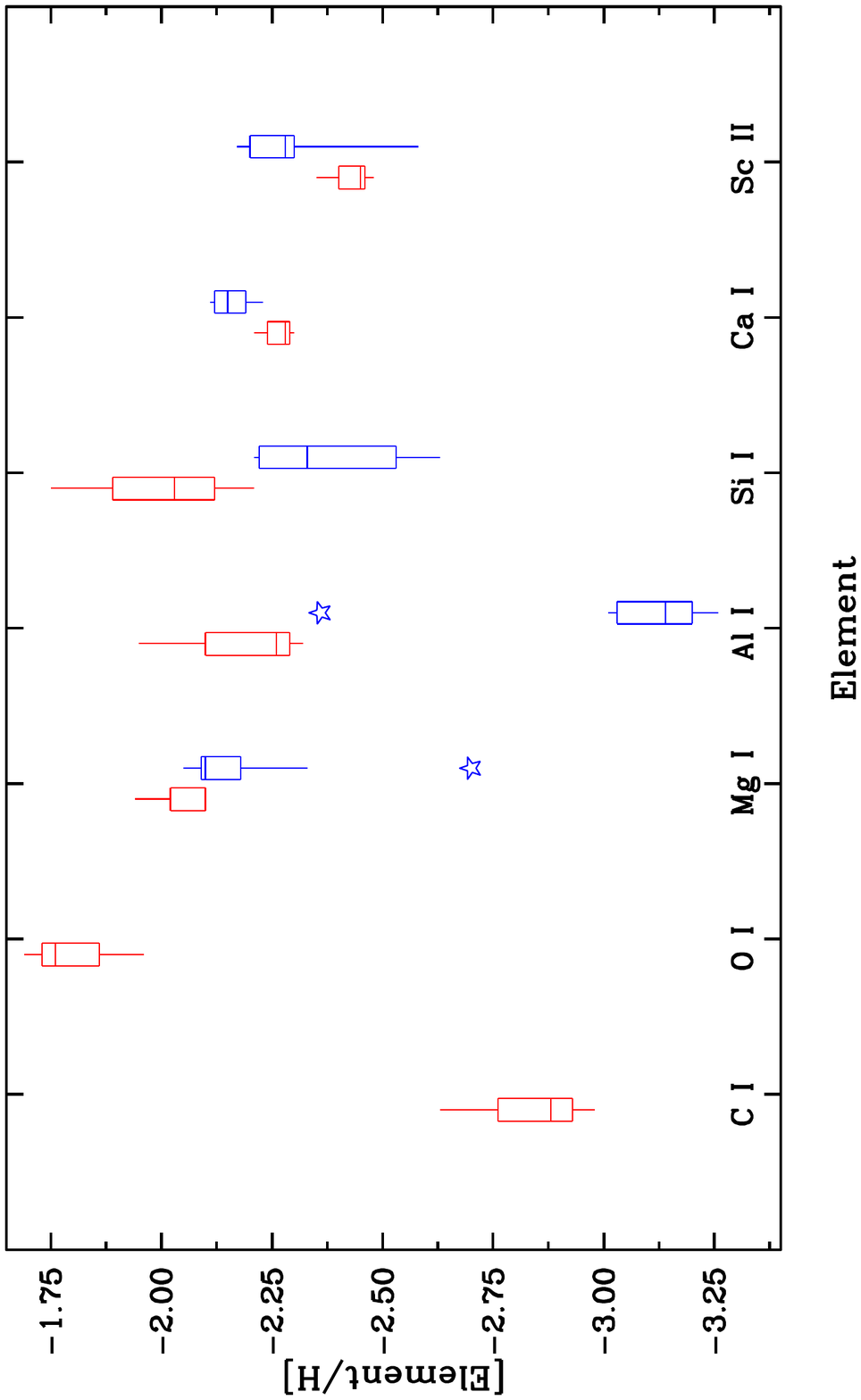}
\caption{
Binned abundances for the elements with $Z = 6, 8, 12-14, 20-21$ displayed as quartile boxplots.  For each species, the boxplot shows the median, 
the upper and lower quartiles, and the extremes for both the RGB (denoted in red) and RHB (signified in blue) samples.  Outliers, data points with values greater than
1.25 times the median, are indicated by star symbols.  Note that there are no RHB points for C I and O I.  Note that a depletion in the relative carbon abundance 
with respect to solar is found. 
\label{light_elem}
}
\end{figure}

\begin{figure}
\centering%
\includegraphics[width=5.9in]{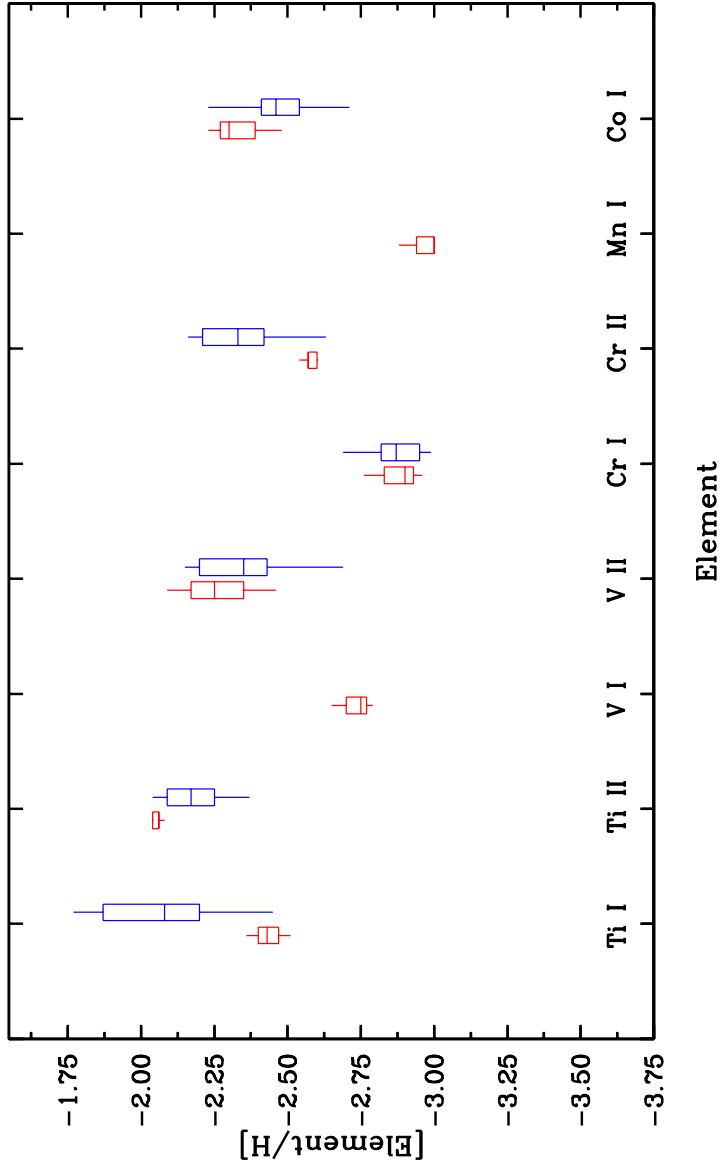}
\caption{
Binned abundances for the elements with $22 \leq Z \leq 27$ displayed as quartile boxplots.  For each species, the boxplot shows the median, 
the upper and lower quartiles, and the extremes for both the RGB (denoted in red) and RHB (signified in blue) samples.  Note that no RHB data exist for V I.  
Also, the Mn I results have been set aside for the RHB stars (consult the text for further information).  The most 
consistent agreement between neutral and first-ionized species is found for Ti in all M15 stars.   
\label{fe_peak}
}
\end{figure}

\begin{figure}
\centering
\includegraphics[width=5.9in]{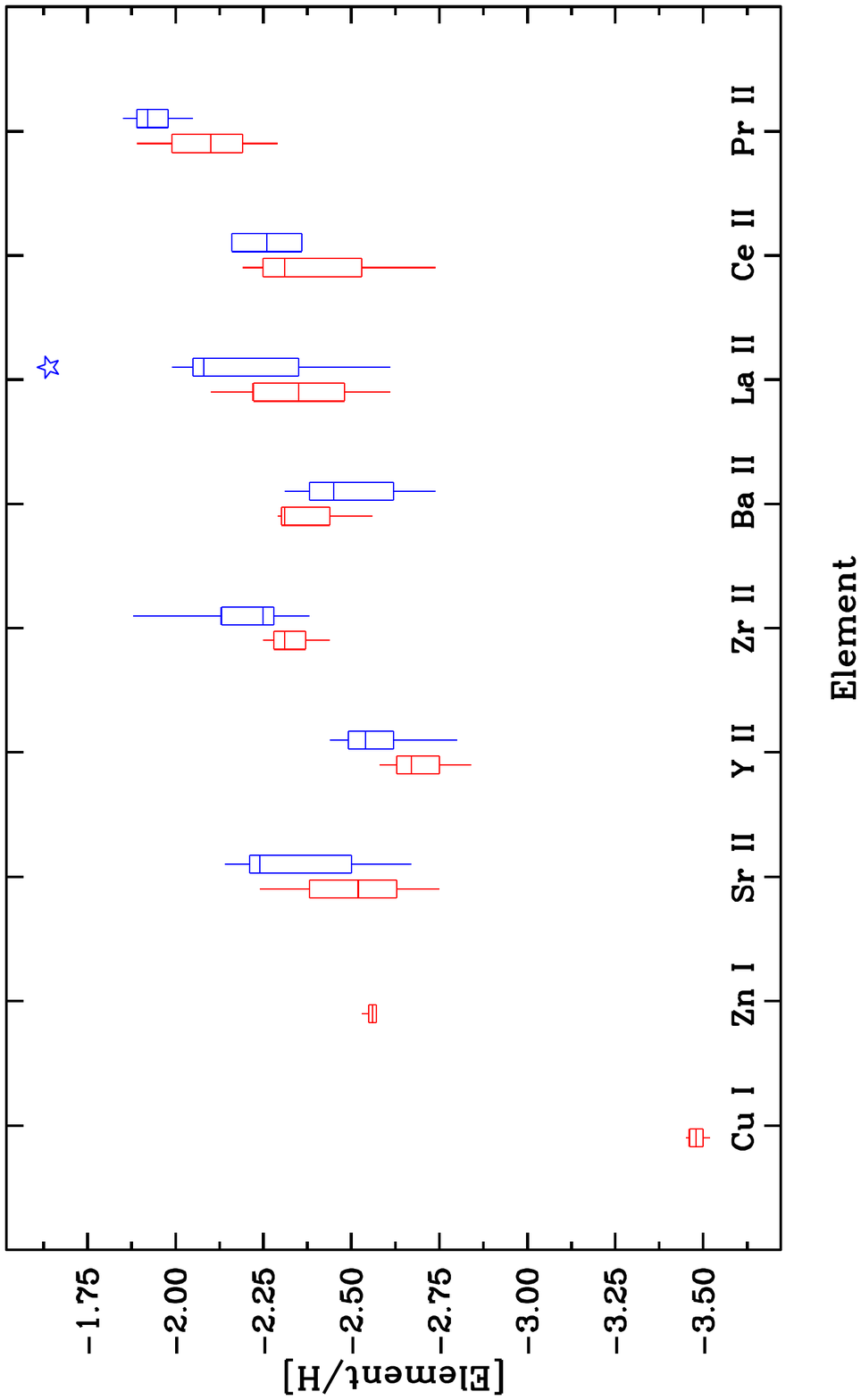}
\caption{
Binned abundances for the elements with $Z = 29, 30, 38-40, 56-59$ displayed as quartile boxplots.  For each species, the boxplot shows the median, 
the upper and lower quartiles, and the extremes for both the RGB (denoted in red) and RHB (signified in blue) samples.  Outliers, data points with values greater than
1.25 times the median, are indicated by star symbols.  Note that there are no RHB data points for Cu I and also, only one RHB abundance value for Zn I.
\label{light_ncapture}
}
\end{figure}

\begin{figure}
\centering%
\includegraphics[width=5.9in]{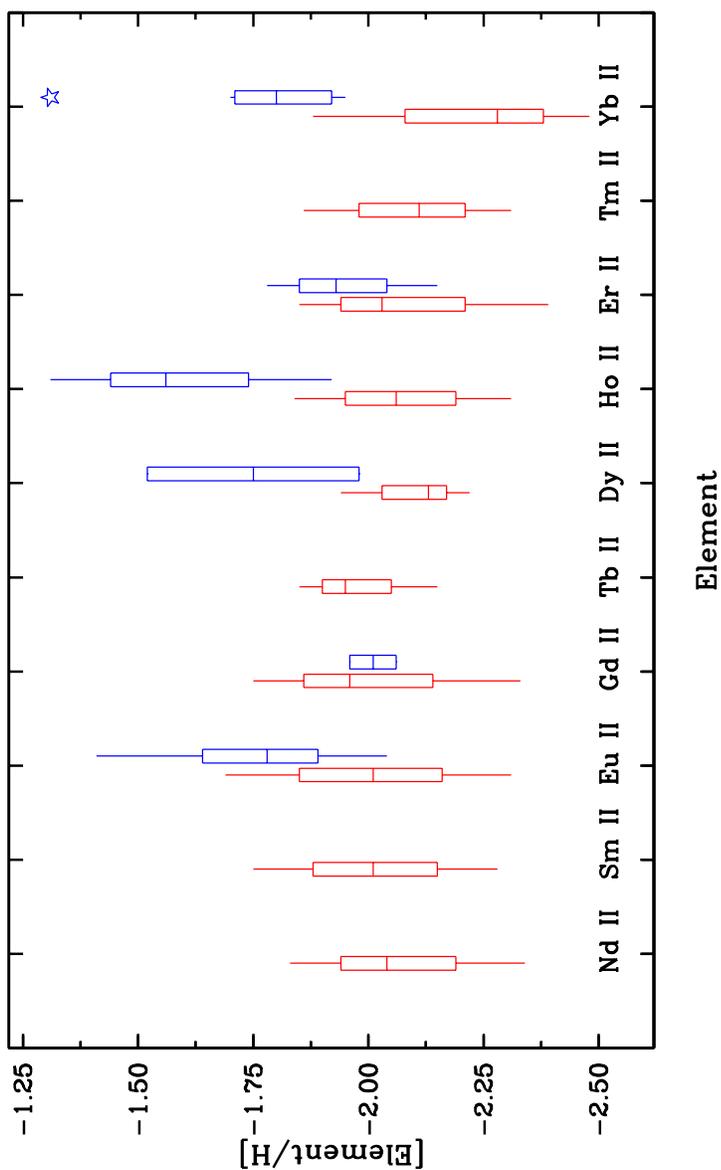}
\caption{
Binned abundances for the elements with $60 \leq Z \leq 70 $ displayed as quartile boxplots.  For each species, the boxplot shows the median, 
the upper and lower quartiles, and the extremes for both the RGB (denoted in red) and RHB (signified in blue) samples.  Outliers, data points with values greater than
1.25 times the median, are indicated by star symbols.  Note that only one RHB abundance value was found for the elements Nd II, Sm II, and 
Tm II.  In the case of Tb II, no RHB measurements were possible. 
\label{heavy_ncapture}
}
\end{figure}

To discuss the abundance results in the following subsections, the elements are divided into four groupings: light ($Z = 8$; $11 \leq Z \leq 21$; 
Figure~\ref{light_elem}), iron-peak ($22 \leq Z \leq 28$; Figure~\ref{fe_peak}), light/intermediate n-capture ($29 \leq Z \leq 59$; Figure~\ref{light_ncapture}), 
and heavy/other ($60 \leq Z \leq 70$; $Z = 72, 76, 77, 90$; Figure~\ref{heavy_ncapture}).  The measurement of abundances was completed for a total of 40 species.  
Note that for the elements Sc, Ti, V, and Cr, abundance determinations were possible for both the neutral and first-ionized species.  In light of 
Saha-Boltzmannn calculations for these elements, greater weight is given to the singly-ionized abundances (i.e. for the stars of the M15 data set, only a small 
fraction of these elements predominantly reside in the neutral state).

Figures~\ref{light_elem},~\ref{fe_peak},~\ref{light_ncapture}, and~\ref{heavy_ncapture} exhibit the abundance ratios for the M15 sample 
in the form of quartile box plots.  These plots show the interquartile range, the median, and the minimum/maximum of the data.  Outliers, points which
have a value greater than 1.25 times the median, are also indicated.  For all of the figures, RGB abundances are signified in red while RHB abundances 
are denoted in blue.  Note that the plots depict the abundance results in the [Elem/H] form in order to preclude erroneous comparisons of the RGB and RHB data, which 
would arise from the iron abundance offset between the two groups.

Table~\ref{m15means} contains the $<$[Elem/Fe]$>$ values for elements analyzed in the M15 sample along with the
line-to-line scatter (given in the form of standard deviations), and the number of lines employed.  The subsequent discussion will generally refer 
to these table data and as is customary, present the {\it relative} element abundances with associated $\sigma$ values.  The reference solar 
photospheric abundances ({\it without non-LTE correction}) are largely taken from the three-dimensional analyses of 
Asplund et al. (2005, 2009\citep{asp05a, asp09}) and Grevesse et al. (2010\citep{gre10}).  However, the photospheric values for some 
of the \ncap\ elements are obtained from other investigations (\eg\ Lawler et al. 2009\citep{law09}).  Table~\ref{solarabund} 
lists all of the chosen log~$\epsilon_{\Sun}$ numbers.  Note that in the derivation of the relative element abundance ratios [X/Fe], $<$[Fe$_{I}$/H]$>$ are employed 
for the neutral species transitions while $<$[Fe$_{II}$/H]$>$ are used for the singly-ionized lines.  This is done in order to minimize ionization 
equilibrium uncertainties as described in detail by Kraft \& Ivans (2003)\citep{kra03}.

\subsection{General Abundance Trends}

Within the RGB sample, the neutron capture element abundances of K462 are consistently the largest whereas those of K583 are the smallest.  The two RHB stars, 
B009 and B262, exhibit abundance trends similar to those of the RGB objects.  The expected anti-correlations in the proton capture elements (\eg\ Na-O and Mg-Al)
are seen.  The greatest abundance variation with regard to the entire M15 data set is found for the neutron capture elements.  Indeed, the star-to-star 
spread for the majority of n-capture abundances is demonstrable for all M15 targets and is not likely due to internal errors.  

Inspection of Table~\ref{m15means} data indicates that RHB stars generally have higher \rpro\ element abundances than RGB stars (on average 
$\Delta$[Elem/Fe]$_{RHB-RGB}\approx $ 0.3 dex).  A sizeable portion of the discrepancy is attributable to the difference in
the iron abundances as $<$[Fe/H]$>_{RHB}$ is approximately 0.12 dex lower than $<$[Fe/H]$>_{RGB}$.  The remaining offset is most likely a consequence of 
the small number of targets coupled with a serious selection effect. The original sample of RHB stars 
from Preston et al. (2006\citep{pre06}) was chosen as a random set of objects with colors and magnitudes representative of the red end of the HB.  These 
objects were selected without prior knowledge of the heavy element abundances.  On the other hand, the three RGB stars from Sneden \etal\ (2000a\citep{sne00a}) 
were particularly chosen as representing the highest and lowest abundances of the \rpro\, as predetermined in the 17 star sample of 
Sneden \etal\ (1997\citep{sne97}). 
 
\subsection{Light Element Abundances}

The finalized set of light element abundances include: C, O, Mg, Al, Si, Ca, and Sc$_{I/II}$.  In general, an enhancement of these element abundances 
relative to solar is seen in the entire M15 data set. 

An underabundance of carbon was found in one RHB and three RGB targets of M15 based on the measurement of CH spectral features.  As the forbidden O I
lines were detectable only in RGB stars, the average abundance ratio for M15 is $<$[O/Fe]$>_{RGB}= +0.75$.  This value is 
substantially larger than that found by Sneden et al. (1997)\citep{sne97}.  A portion of the discrepancy is due to the approximate 0.15 dex 
difference in the $<$[Fe$_I$/H]$>$ values between the two investigations.  The remainder of departure may be attributed to the adoption of 
different solar photospheric oxygen values: the current study employs log~$\epsilon(O)_{\odot}= 8.71$ (Scott et al. 2009\citep{sco09}) 
while Sneden \etal\ uses log~$\epsilon(O)_{\odot}= 8.93$ (Anders \& Grevesse 1989\citep{and89}).  

For the determination of the sodium abundance, the current study relies solely upon the D$_{1}$ resonance transitions.  Table~\ref{m15means} lists the spuriously 
large spreads in the Na abundance for both the RGB and RHB groups. The Na D$_{1}$ lines are affected by the non-LTE phenomenon of {\it resonance} 
scattering (Asplund 2005b\citep{asp05b}; Andrievsky \etal\ 2007\citep{and07}), which MOOG does not take into account.  Also, Sneden \etal\
(2000b\citep{sne00b}) made note of the relative strength and line profile distortions associated with these transitions and chose to discard the [Na/Fe] values 
for stars with \teff\ $>$ 5000 K.  Consequently, the sodium results from the current study are given little weight and are not plotted in Figure~\ref{light_elem}.

Aluminum is remarkable in its discordance: $<$[Al$_{I}$/Fe$_I$]$> = 0.37$ for one RHB and three RGB targets whereas $<$[Al$_{I}$/Fe$_I$]$> = -0.43$ for 
five RHB stars.  Now, the relative aluminum abundances for the RHB stars match well with the values found by Preston \etal\ (2006\citep{pre06}).  
Similarly, the Al abundances from the current analysis agree favorably with the RGB data from Sneden \etal\ (1997\citep{sne97}).  Though relatively strong transitions 
are employed in the abundance derivation, the convergence upon two distinct [Al/Fe] values is nontrivial and could merit further exploration. 

A decidedly consistent Ca abundance ratio is found for the RGB sample: $<[Ca_{I}/Fe_{I}]>_{RGB}= 0.29$; and, also for 
the RHB sample: $<$[Ca$_{I}$/Fe$_{I}$]$>_{RHB}= 0.53$.  After consideration of the iron abundance offset, the RHB 
stars still report slightly higher calcium abundances than the RGB stars.  Overall, a distinct overabundance of Ca relative to solar is present in the M15 cluster.  
Note that the Sc$_{I}$ abundance determination was done for only one M15 star (and gives a rather aberrant result compared to the Sc$_{II}$ abundance
data from the other M15 targets).

\subsection{Iron Peak Element Abundances}

The list of finalized Fe-peak element abundances consists of Ti$_{I/II}$, V$_{I/II}$, Cr$_{I/II}$, Mn, Co, and Ni.  Due to RGB spectral crowding issues, 
derivations of [Ni$_{I}$/Fe$_{I}$] ratios are performed only for RHB stars.

Achievement of ionization equilibrium did not occur for any of the perspective species: Ti$_{I/II}$, V$_{I/II}$, or Cr $_{I/II}$.  In consideration of the
entire M15 data set, the best agreement between neutral and singly-ionized species arises for titanium, with all $<$[Ti/Fe]$>$ ratios being supersolar.  The 
V$_{II}$ relative abundances compare well with one another for the RGB and RHB targets (comparison for V$_{I}$ is not possible as there are no RHB data for
this species).  Both of the RGB and RHB $<$[Cr$_{I}$/Fe$_{I}$]$>$ ratios are underabundant with respect to solar and the neutral 
chromium values match almost exactly with one  another (after accounting for the [Fe/H] offset).  On the other hand, the worst agreement is found for 
Cr$_{I/II}$ in RHB stars with $\Delta(II-I) = 0.47$.

Subsolar values with minimal scatter were found for the $<$[Mn$_{I}$/Fe$_{I}$]$>$ ratios in both the RGB and RHB stellar groups.  However in comparison
to RGB stars, manganese appears to be substantially more deficient in RHB targets.  The discrepancy may be attributed to both the RGB/RHB iron abundance 
disparity as well as the employment of the Mn$_{I}$ resonance transition at 4034.5 \AA\ for the RHB abundance determination.  In particular, 
Sobeck et al. (2011\citep{sob11}) have demonstrated that the manganese resonance triplet (4030.7, 4033.1, and 4034.5 \AA) fails to be a 
reliable indicator of abundance.  Consequently, the RHB abundance results for Mn$_{I}$ are given little weight and are not plotted in Figure~\ref{fe_peak}.

\subsection{Light and Intermediate n-Capture Element Abundances}
  
Finalized abundances for the light and intermediate n-capture elements include Cu, Zn, Sr, Y, Zr, Ba, La, Ce, and Pr.  In general, the RGB element abundance ratios 
are slightly deficient with respect to the RHB values.  Also, enhancement with respect to solar is consistently seen in all M15 targets for the elements Ce and Pr.   

An extremely underabundant copper abundance relative to solar was found in the RGB stars: $<$[Cu$_{I}$/Fe$_{I}$]$>_{RGB} = -0.91$.  A similar 
derivation could not take place in the RGB targets as the Cu$_I$ transitions were too weak.  A large divergence between RGB and RHB stellar 
abundances exists for zinc.  Detection of the Zn transitions was possible in only one RHB target, which could perhaps account for some of the discrepancy.  

For the entire M15 data set, Y$_{II}$ exhibits lower relative abundance ratios in juxtaposition to both Sr$_{II}$ and Zr$_{II}$.  With regard to the three 
average elemental abundances (of Sr, Y, Zr), moderate departures between the RGB and RHB groups are seen.  Also, a large variation in 
the $<$[Sr$_{II}$/Fe$_{II}$]$>$ ratio was found for the members of the RGB group. 

Though different sets of lines are employed, the RGB and RHB $<$[Ba$_{II}$/Fe$_{II}$]$>$ ratios are consistent with one another.  A portion of the RHB 
abundance variation is due to the exclusive use of the resonance transitions in the determination (two lowest temperature RHB stars 
report quite high $\sigma$ values; these strong lines could not be exploited in the RGB analysis).  Notably for this element group, the greatest star-to-star 
abundance scatter was found for lanthanum: $\Delta_{RGB} = 0.46$ and $\Delta_{RHB} = 0.61$ (excluding the one RHB outlier).  The 
relative cerium abundances also exhibit a wide spread in the RGB sample.

\subsection{Heavy n-Capture Element Abundances}

The list of finalized $<$[El/Fe]$>$ ratios for the heavy n-capture elements is as follows: Nd, Sm, Eu, Gd, Tb, Dy, Ho, Er, Tm, Yb, Hf, 
Os, Ir, Pb and Th.  All of these element abundance ratios are enriched with regard to the solar values.  As shown in 
Figure~\ref{heavy_ncapture}, a larger abundance spread is found for this group in comparison to the other element groups.

Note that as \teff\ increases, the strength of the heavy element transitions rapidly decreases and as a consequence, the use of these lines 
for abundance determinations in the warmest stars becomes unfeasible.  It was possible to obtain robust abundances for Nd, Sm, Tb, and Tm in a single 
RHB target.  On the other hand, abundance extractions for the species Os and Ir were done only in RHB stars (measurements of these element transitions 
were attainable as less spectral crowding occurs in these stars).  Nonetheless, minimal line-to-line scatter is seen for the bulk of RGB and RHB n-capture abundances.

A rigorous determination of the europium relative abundance was performed for all M15 stars: $<$[Eu$_{II}$/Fe$_{II}$]$>_{RGB} = 0.53$ and 
$<$[Eu$_{II}$/Fe$_{II}$]$>_{RHB} = 0.88$.  Despite the iron abundance offset, the largest departure between the two stellar groups is found for 
the element Ho$_{II}$.  Further, the greatest star-to-star scatter in the heavy n-capture elements is seen for the [Yb$_{II}$/Fe$_{II}$] ratio: $\Delta_{RGB} = 0.55 $ and 
$\Delta_{RHB} = 0.58$.

\subsection{Comparison with Previous CTG efforts and Otsuki \etal\ (2006)}\label{otsuki}

These new abundance results are now compared to those from the four prior CTG publications.  For the majority of elements, the current data are in
accord with the findings of Sneden et al. (1997, 2000a\citep{sne97,sne00a}) and Preston et al. (2006\citep{pre06}).  In this 
effort, abundance derivations are performed for 13 new species: Sc$_{I}$, V$_{II}$, Cu$_{I}$, 
Pr$_{II}$, Tb$_{II}$, Ho$_{II}$, Er$_{II}$, Tm$_{II}$, Yb$_{II}$, Hf$_{II}$, Os$_{I}$, Ir$_{I}$, and Pb$_{II}$.  For elements
re-analyzed in the current study, the abundance data have been improved with the use of higher quality atomic data, additional transitions, and a revised version of the 
MOOG program.  A few large discrepancies in the [El/Fe] ratios do occur between the current study and the previous M15 efforts.  These departures can be 
attributed to the employment of different [Fe/H] and solar photospheric values as well as the updated MOOG code.  Accordingly, the results from the 
current analaysis supersede those from the earlier CTG papers.

As in Sneden \etal\ 1997\citep{sne97}, the abundance behavior of the proton capture elements appears to be {\it decoupled} from 
that of the neutron capture elements.  Notably for M15, significant spread in the abundances was confirmed for both Ba and Eu.  The scatter  
of $\Delta$log$\epsilon(Ba)= 0.48$ and $\Delta$log$\epsilon(Eu)= 0.90$ from the current effort is in line with 
that of $\Delta$log$\epsilon(Ba)= 0.60$ and $\Delta$log$\epsilon(Eu)= 0.73$ from Sneden \etal\ (1997)\citep{sne97}.

Comparison of the findings from the current study to those from Otsuki \etal\ (2006)\citep{ots06} has also been done and will be 
limited to the only star that the two investigations have in common, K462.  Due to differences in the $<$[Fe/H]$>$ values, the 
log~$(\epsilon)$ data of the two analyses are compared.  The model atmospheric parameters for K462 differ somewhat between the 
current effort (\teff/\logg/\vturb\ = 4400/0.30/2.00) and Otsuki \etal\ (\teff/\logg/\vturb\ = 4225/0.50/2.25).  However, the agreement
in the abundances for the elements Y, Zr, Ba, La and Eu is rather good between the two studies, with the exact differences ranging: 
$0.01 \leq \mid\Delta(Otsuki-Current)\mid \leq 0.16$.  The largest disparity occurs for Sr, with both analyses employing the resonance transitions.  As
mentioned previously, these lines are not the most rigorous probes of abundance.

\subsection{General Relationship of Ba, La, and Eu Abundances}

\begin{figure}
\centering%
\includegraphics[width=5.9in]{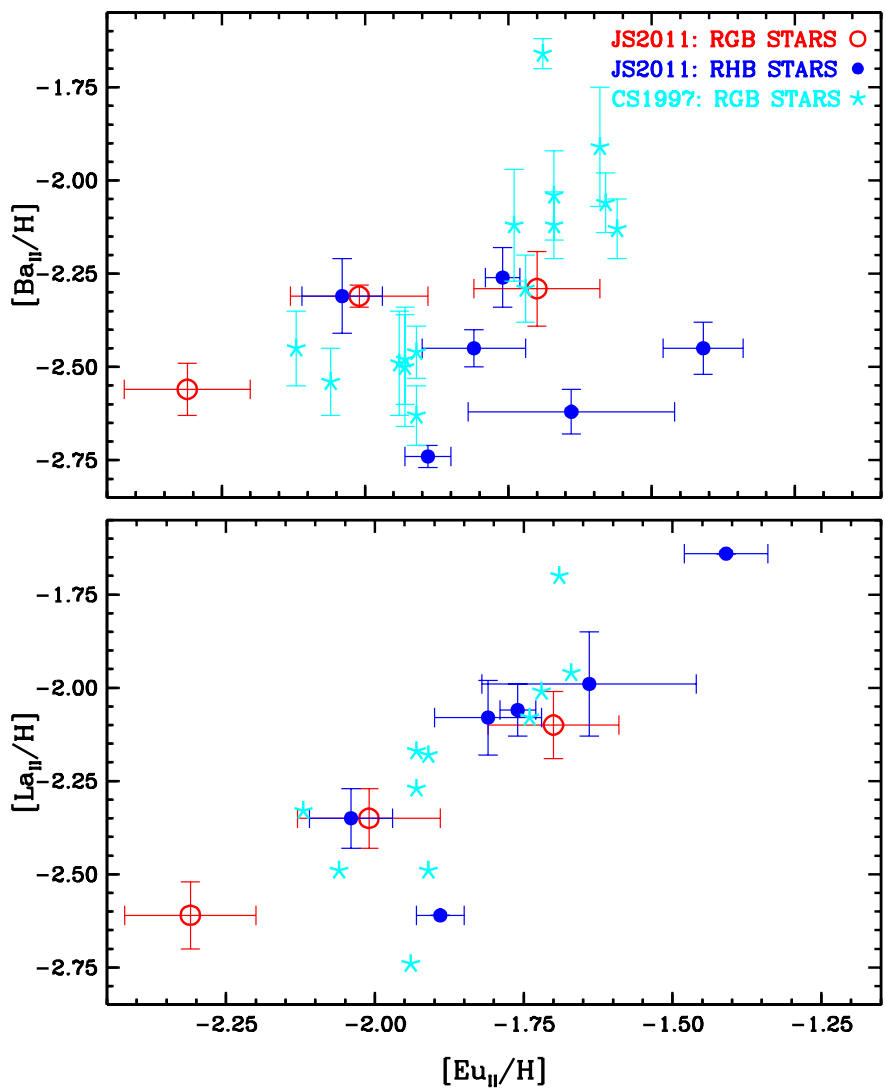}
\caption{
Plots of [(Ba, La)/H] as a function of [Eu/H] for the M15 stars of the current study as well as those from the original Sneden et al. (1997) paper.  Also shown
are the associated error bars for the abundances (these are the standard deviation values given in Table~\ref{m15means}).  Note that the 
Eu and La abundance determinations for the 1997 stellar sample are based upon one transition only (and consequently, no error bars are plotted).  Contrary 
to Sneden et al., no clear evidence of a binary distribution in the abundances is detected.  
\label{Ba_La_Eu}
}
\end{figure}

Sneden et al. (1997)\citep{sne97} claimed to have found a binary distribution in a plot of [Ba/Fe] versus [Eu/Fe], with 8 stars exhibiting
relative Ba and Eu abundances approximately 0.35 dex smaller than the remainder of the M15 data set.  To re-examine their assertion, 
Figure~\ref{Ba_La_Eu} is generated, which plots [(Ba, La)/H] as a function of [Eu/H] for the entire data sample of the current study.  It also displays 
the re-derived/re-scaled Ba, La, and Eu abundances for all of the giants from the Sneden et al. (1997)\citep{sne97} publication.  No decisive offset is 
evident in either panel of Figure~\ref{Ba_La_Eu}.  For completeness, the equivalent width data from Otsuki et al. (2006)\citep{ots06} were 
also re-analyzed and the abundances were re-determined.  Again, no bifurcation was detected in the Ba and Eu data.\footnote{To avoid duplication, 
the stars from Ostuki et al. are not plotted as they are a subset of the original sample from the Sneden et al. (1997\citep{sne97}) study.}  

\section{DATA INTERPRETATION AND ANALYSIS}

\begin{figure}
\includegraphics[width=5.4in]{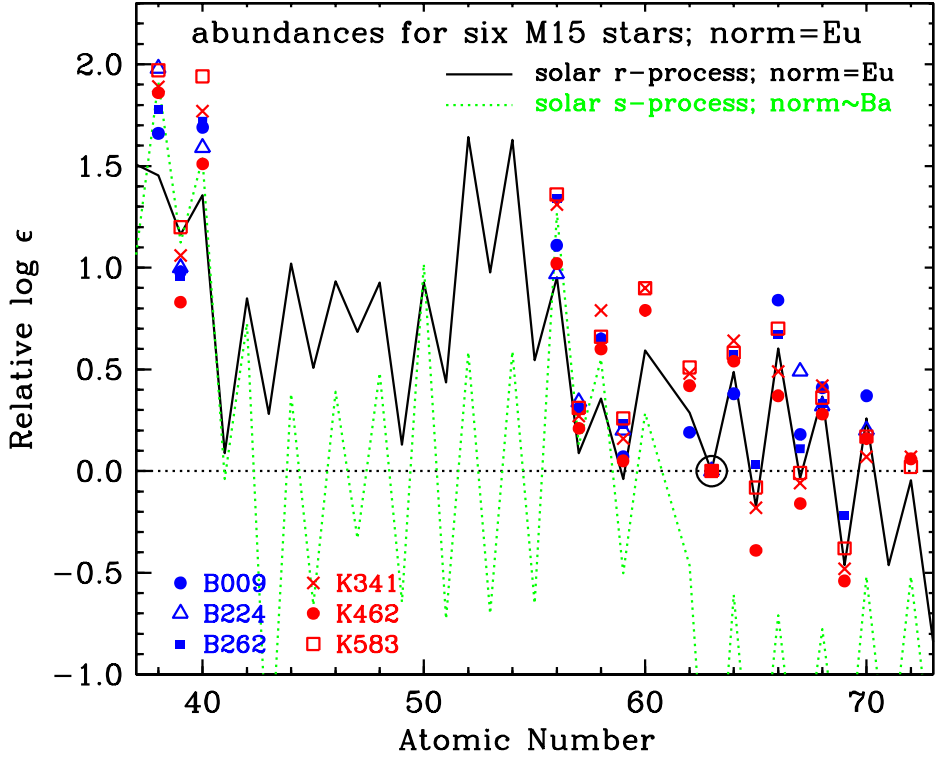}
\caption{
Comparison of the n-capture abundances of six M15 targets to the Solar System r- and s-only abundance distributions.  The abundances of the three RGB stars 
(K341, K462, and K583) are signified by red symbols while the abundances of the three RHB stars (B009, B224, and B262) are designated by 
blue symbols.  The solid black line denotes the \rpro\ only abundance pattern, which is scaled to the solar log~$\epsilon(Eu)$, and 
the dotted green line indicates the \spro\ only abundance pattern, which is scaled to the solar log~$\epsilon(Ba)$ (all predictions are taken 
from Sneden, Cowan, \& Gallino 2008).  The heavy element abundances of the M15 stars compare well to the \rpro\ predictions, but not those of the \spro.  Neither 
abundance distribution consistently matches the stellar abundances for the elements Sr, Y, and Zr.
\label{abundance_Z}
}
\end{figure}

A significant amount of \rpro\ enrichment has occurred in the M15 globular cluster.  Figure~\ref{abundance_Z} plots
the average log~$\epsilon$ values of the n-capture elements (with $39 \leq Z \leq 70$) for three RGB stars (K341, K462, K583; signified by red symbols) 
and three RHB stars (B009, B224, B262; denoted by blue symbols).\footnote{B028, B412, and B584 are not included in 
the figure as these stars lack abundances for most of the elements in the specified Z range.}  The solid black line in this figure indicates the scaled, solar 
\rpro\ prediction as computed by Sneden, Cowan, \&\ Gallino (2008)\citep{sne08}.  All of the element abundances 
are normalized to the individual stellar log~$(\epsilon_{Eu})$ values (Eu is assumed to be an indicator of \rpro\ contribution).  For the n-capture 
elements with $Z = 64-72$, the RGB stellar abundance values {\it strongly correlate} with the solar \rpro\ distribution.  Similarly for the RHB stars, 
these abundances match well to the solar \rpro\ pattern for most of the elements in the $Z = 64-72$ range.  

Figure~\ref{abundance_Z} also displays the scaled, solar \spro\ abundance distribution (green, dotted line). The \spro\ predictions are also taken from
Sneden, Cowan, \& Gallino and the values are normalized to the solar log~$(\epsilon_{Ba})$ (Ba is considered to be an indicator of \spro\ contribution).  
As shown for the $Z = 64-72$ elements, there is {\it virtually no} agreement between the solar \spro\ pattern and either the RGB or the RHB stellar abundances.  The 
\spro\ predictions compare well to the RGB abundances for only two elements: Ce and La.  Thus, it follows that the nucleosynthesis 
of the {\it heavy} neutron capture elements in M15 was dominated by the \rpro.  In addition, the abundance pattern for the light n-capture elements (Sr, Y, Zr) 
does not adhere to either a solar \rpro\ or \spro\ distribution.

\subsection{Evidence for Additional Nucleosynthetic Mechanisms Beyond the Classical {\it r-} and \spro\ }

\begin{figure}
\centering%
\includegraphics[width=5.9in]{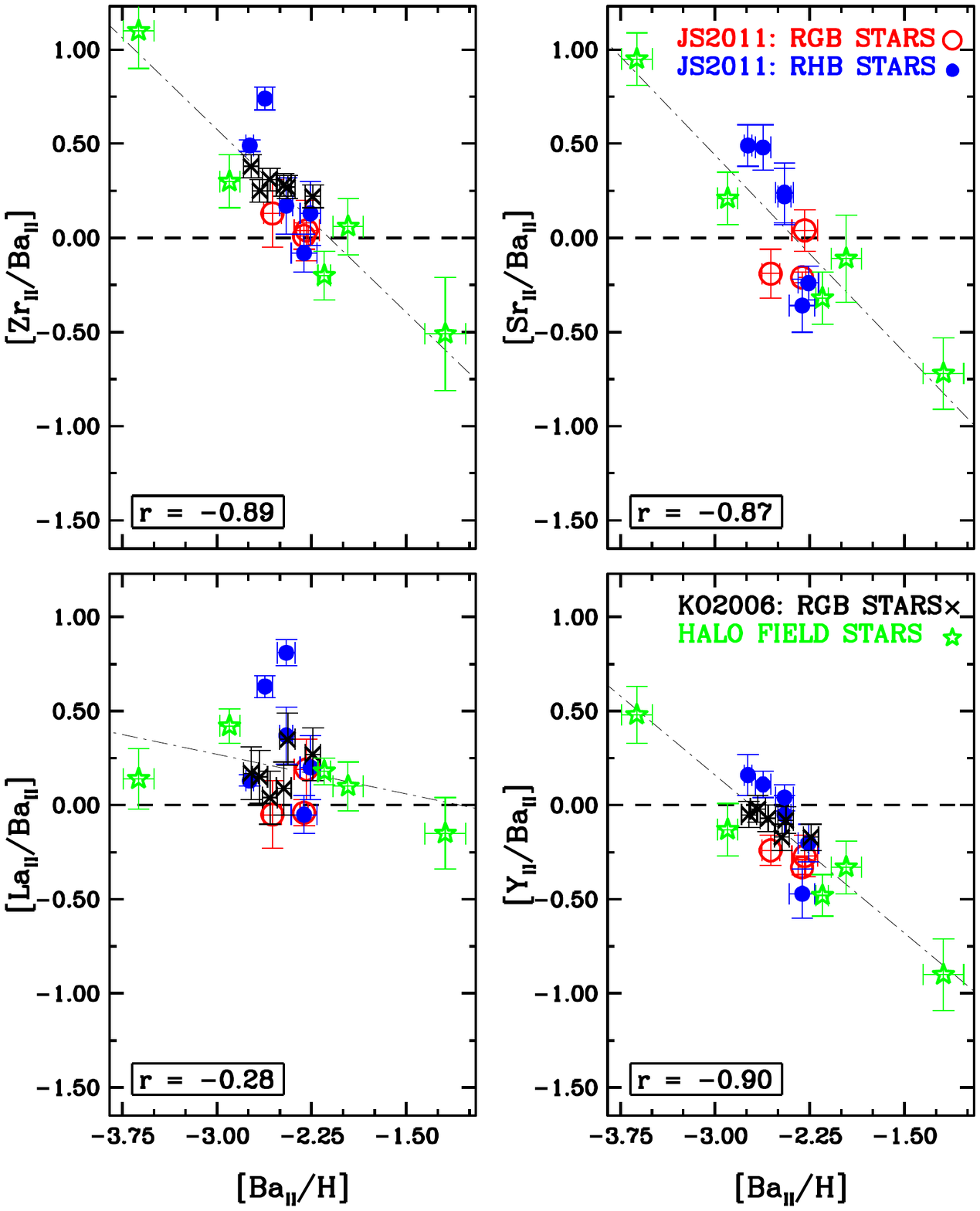}
\caption{
Plot of the relative abundances of the light n-capture elements Sr, Y, and Zr as function of the traditional \spro\ indicator, Ba.  All M15 stars from
the current effort are shown.  Also displayed are the abundance results from Otsuki et al. (2006) as well as those for a few halo field stars (taken
from various literature references; see text for further details).  The correlation coefficient, {\bf r}, is exhbited in each of the panels and as
seen, Sr, Y , and Zr demonstrate a clear anti-correlative behavior with Ba while La does not.
\label{SrYZr_Ba}
}
\end{figure}

\begin{figure}
\centering%
\includegraphics[width=5.9in]{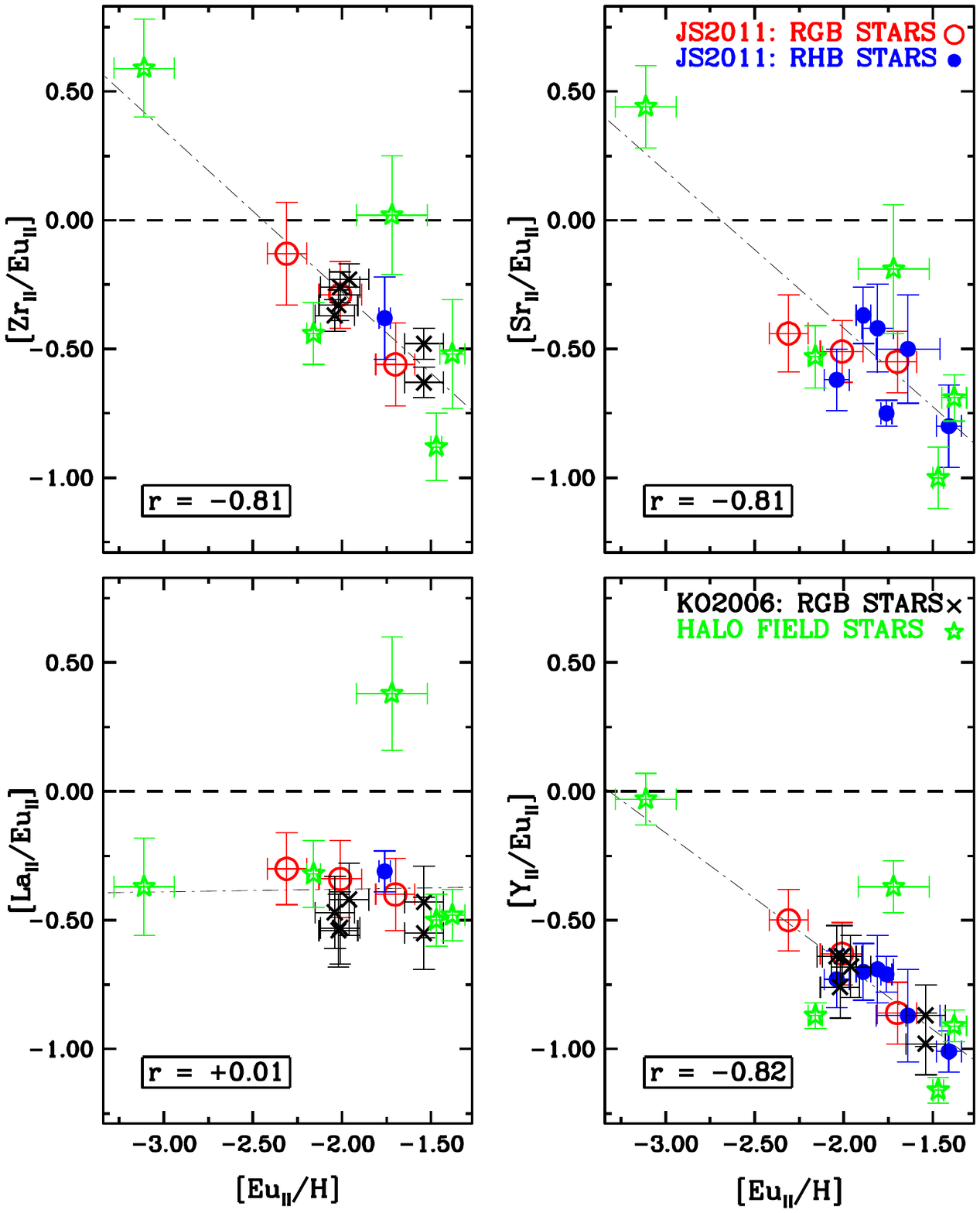}
\caption{Plot of the relative abundances of the light n-capture elements Sr, Y, and Zr as function of the traditional \rpro\ indicator, Eu.  All M15 stars from
the current effort are shown.  Also displayed are the abundance results from Otsuki et al. (2006) as well as those for a few halo field stars (taken
from various literature references; see text for further details).  The correlation coefficient, {\bf r}, is exhbited in each of the panels and as
seen, Sr, Y , and Zr demonstrate a clear anti-correlative behavior with Eu while La does not.
\label{SrYZr_Eu}
}
\end{figure}

To further examine the anomalous light n-capture abundances in the M15 cluster, Figures~\ref{SrYZr_Ba} and \ref{SrYZr_Eu} are generated.  For the
stars of current effort and those from Otsuki et al. (2006)\citep{ots06}, these two plots display the abundances of the n-capture elements (Sr, Y, Zr and La) 
as a function of the [Ba/H] and [Eu/H] ratios, respectively.  Moreover, the abundance results from five select 
field stars which represent extremes in \rpro\ or \spro\ enhancement are plotted (CS 22892-052 [Sneden et al. 2003, 2009\citep{sne03,sne09}]; CS 22964-161 
[Thompson et al. 2008]\citep{tho08}; HD 115444 [Westin et al. 2000\citep{wes00}, Sneden et al. 2009\citep{sne09}]; HD 122563 [Cowan et al. 2005\citep{cow05}, 
Lai et al. 2007\citep{lai07}]; HD 221170 [Ivans et al. 2006\citep{iva06}, Sneden et al. 2009\citep{sne09}]).

In Figure~\ref{SrYZr_Ba}, an anti-correlative trend is seen for Sr, Y and Zr with Ba while no 
explicit correlative behavior is apparent for La.  The correlation coefficient, $r$, is indicated in each panel.  Likewise, La and Eu appear 
un-correlated in Figure \ref{SrYZr_Eu}.  The [(Sr, Y, Zr)/Eu] ratios all exhibit anti-correlation with [Eu/H] in this figure.  As shown, the elements Sr, Y, and Zr 
clearly demonstrate an anti-correlative relationship with {\it both} the markers of the \spro\ (Ba) and the \rpro\ (Eu). 

Figures~\ref{SrYZr_Ba} and~\ref{SrYZr_Eu} collectively imply that the production of the light neutron capture elements most likely did not transpire 
via the classical forms of the \spro\ or the \rpro.  This finding is not novel.  The abundance survey of halo field stars by Travaglio et al. (2004)\citep{tra04} 
previously established the decoupled behavior of the light n-capture species to both Ba and Eu.  Further, they postulated that an additional nucleosynthetic
process was necessary for the production of these elements (Sr, Y , Zr) in metal-deficient regimes (coined the Lighter Element Primary Process; LEPP).  

\begin{figure}
\centering%
\includegraphics[width=5.30in]{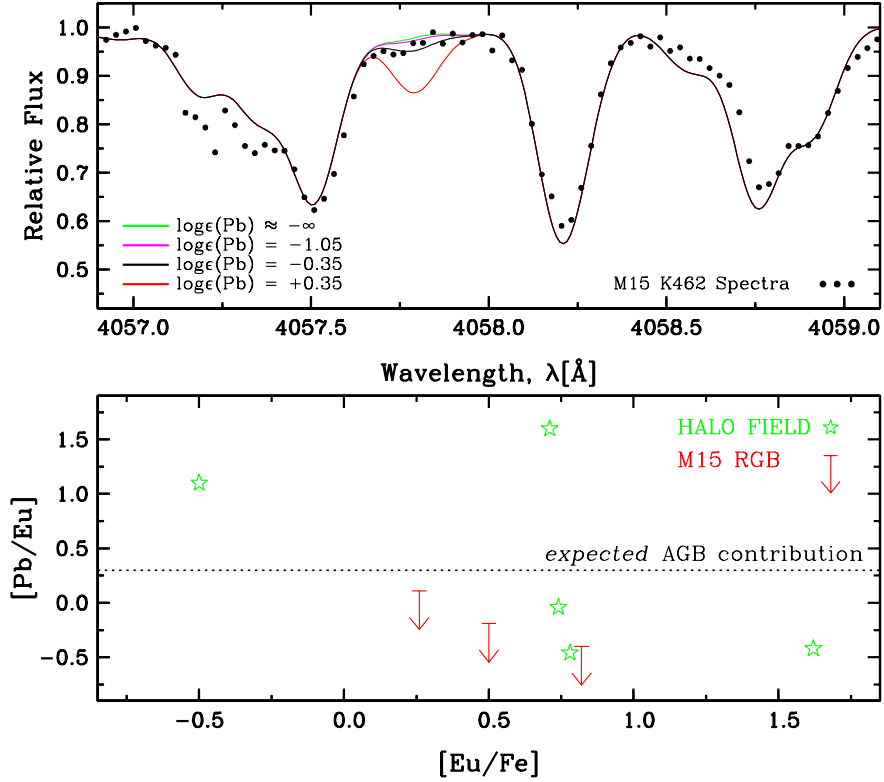}
\caption{{\it Upper panel:} Comparison of synthetic and observed spectra for the Pb I transition at 4057.8 \AA\ in the M15 K462 giant.  Four incremental changes in 
abundance are shown for the specified Pb feature.  It is only possible to establish an upper limit of approximately log~$\epsilon$ (Pb)$\approx$ -0.35 for this star.  
A CH contaminant is present in the blue wing of the Pb transition and accounts for a definitive portion of the signal.  Accordingly, it appears that the \spro\ 
element Pb is nominally detected in K462.  {\it Lower Panel:} Plot of [Pb/Eu] as a function of [Eu/Fe] for three M15 giants and 5 halo field stars.  The empirically 
determined threshold ratio, which indicates the occurrence of AGB enrichment, is shown by the short-dashed line at [Pb/Eu] $\geq +$0.3 
(this value is taken from Roederer et al. 2010; see the text for further details).  Notice that all of the M15 stars fall below this line 
(and correspondingly, should lack main \spro\ material).
\label{Pb_sprocess}
}
\end{figure}

The overabundances of Sr and Zr (see Figure~\ref{abundance_Z}) could have been the result of a small \spro\ contribution to 
the M15 proto-cluster environment.  To investigate this possibility, an abundance determination is performed for Pb, a definitive main \spro\ product.  
The upper panel of Figure~\ref{Pb_sprocess} illustrates the synthetic spectrum fits to the neutral Pb transition at a wavelength of 4057.8 \AA\
in the M15 giant, K462.  An upper limit of log~$\epsilon(Pb) \lesssim$ -0.35 can only be established for this star.  For the remaining two RGB targets, 
upper limits were also determined and accordingly for all three, the average values of log~$\epsilon(Pb) \lesssim$~-0.4 and $<$[Pb/Eu]$>$ $\lesssim$~-0.15 were found.

The lower panel of Figure~\ref{Pb_sprocess} plots [Pb/Eu] as a function of [Eu/Fe] for the three M15 RGB stars and the five, previously-employed halo 
field stars.  In a recent paper, Roederer \etal\ (2010\citep{roe10}) suggest that detections of Pb and enhanced [Pb/Eu] ratios should be strong indicators 
of main \spro\ nucleosynthesis.  In turn, they contend that non-detections of Pb and depleted [Pb/Eu] ratios should signify the absence of nucleosynthetic 
input from the main component of the \spro\ (see their paper for further discussion).  With the abundances of 161 low-metallicity stars 
([Fe/H] $<$ -1), Roederer \etal\ empirically determined a threshold value of [Pb/Eu] = +0.3 for minimum AGB contribution.  As shown in the figure, 
the M15 giants lie below this threshold and accordingly, are likely devoid of main \spro\ input. Thus in the case of the M15 globular cluster, the light neutron capture 
elements presumably originated from an alternate nucleosynthetic process (\eg\ $\nu-p$ process, Frohlich \etal\ 2006\citep{fro06}; high entropy winds, 
Farouqi \etal\ 2009\citep{far09}).   

\subsection{M15 Abundances in Relation to the Halo Field}

\begin{figure}
\centering%
\includegraphics[width=5.7in]{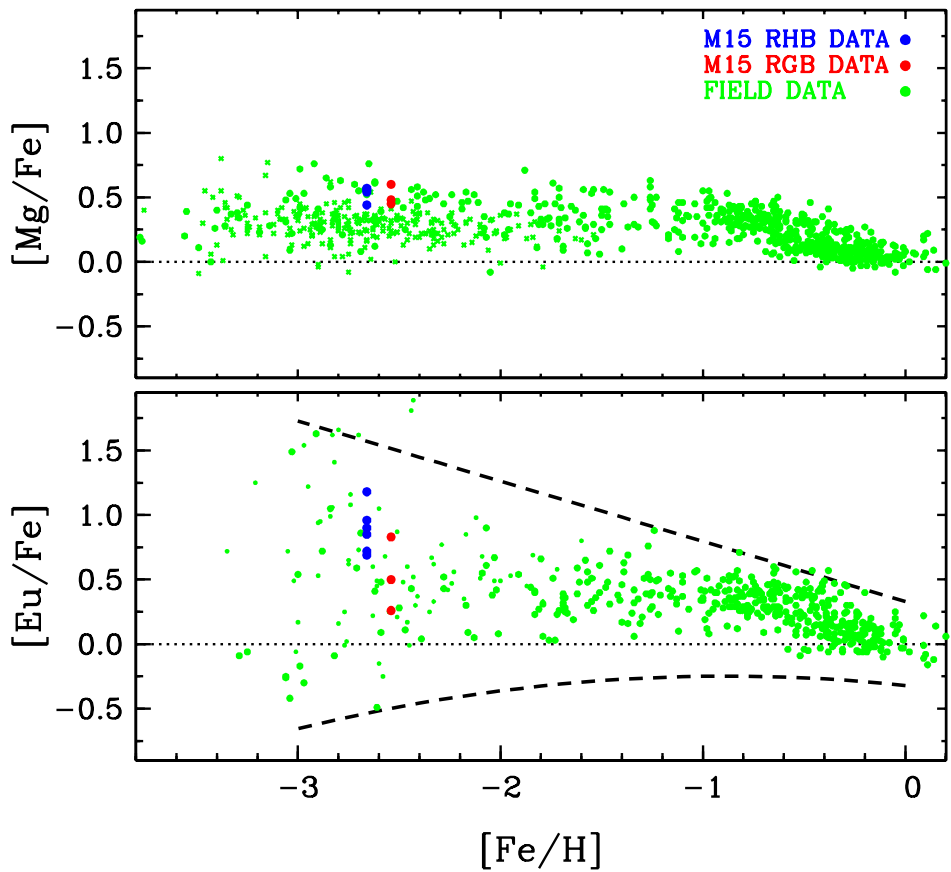}
\caption{
Comparison of the distribution [(Mg, Eu)/Fe] relative abundances as a function of metallicity for M15 targets (denoted by the blue and red circles) as well as for halo and
disk stars (signified by the green, filled circles).  Field star data have been taken from large sample surveys (see the text for further information).  As per 
convention, the dotted lines represent the Solar abundance ratios. In in the lower panel, the two dashed black lines indicate the approximate ranges of the [Eu/Fe] data.  
The spread in the [Mg/Fe] values is smaller for M15 than that for the field.  Yet for [Eu/Fe], the M15 scatter is {\it comparable} to
that of the field (for that particular metallicity).
\label{Eu_spread}
}
\end{figure}

The upper panel of Figure~\ref{Eu_spread} displays the evolution of the [Mg/Fe] abundance ratio with [Fe/H] for all M15 targets 
as well as for a sample of hundreds of field stars.  For this figure, halo and disk star data have been taken from these surveys: Fulbright (2000)\citep{ful00}, 
Reddy et al. (2003)\citep{red03}, Cayrel et al. (2004)\citep{cay04}, Cohen et al. (2004)\citep{coh04}, Simmerer et al. (2004)\citep{sim04}, 
Barklem et al. (2005)\citep{bar05}, Reddy, Lambert \& Allende Prieto (2006)\citep{red06}, Fran{\c c}ois et al. (2007)\citep{fra07}, 
and Lai et al. (2008)\citep{lai08}.  As shown, the scatter in the [Mg/Fe] abundance ratio is fairly small: $\Delta ($[Mg/Fe]$)_{MAX} \approx 0.6$ dex 
for all stars under conisderation and $\Delta ($[Mg/Fe]$)_{MAX} \approx 0.1$ dex for the M15 data set.  In the metallicity regime
below [Fe/H] $\lesssim$ -1.1, the roughly consistent trend of [Mg/Fe] abundance ratio is due in part to the production history for
these elements: magnesium originates from hydrostatic burning in massive stars while iron is manufactured by massive star, core-collapse SNe.  
If the short evolutionary lifetimes of these massive stars are taken into context with the abundance data,
it would seem to indicate that the core-collapse SNe are rather ubiquitous events in the Galactic halo.  Accordingly, the 
products that result from both stellar and explosive nucleosynthesis of massive stars should be well-mixed in the interstellar and 
intercluster medium.  The apparent downward trend in the [Mg/Fe] ratio, in the metallicity region with [Fe/H]$\gtrsim$ -1.1, 
is due to nucleosynthetic input from Type Ia SNe, which produce much more iron in comparison to Type II events. 

In a similar vein, the lower panel of Figure~\ref{Eu_spread} plots [Eu/Fe] as a function of [Fe/H] and demonstrates that as the metallicity decreases, the spread
in the [Eu/Fe] abundance ratio increases enormously\footnote{Though the data sample of Figure~\ref{Eu_spread} is compilation of several sources, the scatter in the [Mg/Fe] 
and [Eu/Fe] ratios duplicates that found by such large scale surveys as, \eg\, Barklem et al. (2005)}.  By contrast, the scatter in the M15 [Eu/Fe] ratios is large and 
comparable to the spread of the halo field at that metallicity.  Specifically in the metallicity interval -2.7 $\leq$ [Fe/H] $\leq$ -2.2, the scatter in the 
[Eu/Fe] ratio is found to be $\sigma = \pm 0.27$ for the nine stars of the M15 sample and similarly for the 23 halo giants, the
associated scatter is $\sigma = \pm 0.33$.  This variation in the relative europium abundance ratio (as first detected by Gilroy \etal\ 1988\citep{gil88} 
and later confirmed by others, e.g. Burris et al. 2000\citep{bur00}, Barklem et al. (2005)\citep{bar05}) indicates an inhomogeneous production history for Eu and other 
corresponding \rpro\ elements.  These elements likely originate from lower mass SNe and their production is not correlated 
with that of the alpha elements (Cowan \& Thielemann 2004\citep{cow04}).  Furthermore, it seems that nucleosynthetic events 
which generated the \rpro\ elements were rare occurrences in the early Galaxy.  As a consequence, these elements 
were not well-mixed in the interstellar and intercluster medium (Sneden et al. 2009\citep{sne09}).  Note that \rpro\ enhancement seems to be a common feature of all 
globular clusters (e.g. Gratton, Sneden, \& Carretta 2004\citep{gra04}).  On the other hand, the scatter in select \rpro\ element abundances, as found in M15, 
is {\it not}.

\section{SUMMARY}

A novel effort was undertaken to perform a homogenous abundance determination in {\it both} the RGB and RHB members of the M15 globular cluster.  The current 
investigation employed improved atomic data, stellar model atmospheres, and radiative transfer code.  A resolute offset in the 
iron abundance between the RGB and RHB stars on the order of 0.1 dex was measured.  Notwithstanding, the major findings of the analysis for {\it both} 
the RGB and RHB stellar groups include: a definitive \rpro\ enhancement; a significant spread in the abundances of the neutron capture species (which appears
to be astrophysical in nature); and, an anti-correlation of light n-capture element abundance behavior with both barium ([Ba/H]) and europium ([Eu/H]).  Accordingly, 
the last set of findings may offer proof of the operation of a LEPP-type mechanism within M15.  To determine if these abundance behaviors are generally 
indicative of {\it very} metal-poor globular clusters, a comprehensive examination of the chemical composition of the analogous M92 cluster should be undertaken
([Fe/H] $\sim$ -2.3, Harris et al. 1996\citep{har96}; the literature contains relatively little information with regard to the 
n-capture abundances for this cluster). 

To date, the presence of multiple stellar generations within the globular cluster M15 has not been irrefutably established.  In a series of papers, 
Carretta et al. (2009a, 2009c, 2010\citep{car09a,car09c,car10}) offered compelling proof in the detection of light element anti-correlative behavior (Na-O) 
in numerous members of the M15 RGB.  Lardo et al. (2011\citep{lar11}) did find a statistically significant spread in the SDSS photometric 
color index of $u-g$, but yet they were not able to demonstrate a clear and unambiguous correlation of $(u-g)$ with the Na abundances in the RGB 
of the M15 cluster (which would have provided further evidence).  To wit, recent investigations of M15 have revealed several atypical features 
including: probable detection of an intermediate mass black hole (van der Marel et al. 2002\citep{van02}; though the result is under some dispute); 
observation of an intracluster medium (Evans et al. 2003\citep{eva03}); detection of mass loss (Meszaros et al. 2008, 2009\citep{mez08,mez09}); 
identification of extreme horizontal branch and blue hook stars (Haurberg \etal\ 2010\citep{hau10}); and, observation of an 
extended tidal tail (Chun et al. 2010\citep{Chun2010}).  It would be worthwhile to examine these peculiar aspects of the globular cluster 
in relation to the abundance results of M15.  Further scrutiny is warranted in order to understand the star formation history and mixing timescale 
of the M15 protocluster environment.  

\acknowledgments 

We are deeply indebted to L. Koesterke for his extensive advice with regard to the modification of the MOOG code.  We are grateful to the referee for several 
valuable suggestions.  We also thank I. Roederer for helpful comments pertaining to drafts of the manuscript. The current effort has made use of the 
NASA Astrophysics Data System (ADS), the NIST Atomic Spectra Database (ASD), and the Vienna Atomic Line Database (VALD).  Funding for this research has been 
generously provided by the National Science Foundation (grants AST 07-07447 to J.C. and AST 09-08978 to C.S.).  
\clearpage

\appendix 
\label{Ap}

\section{Synopsis}

The essential approach to the solution of radiative transfer in the MOOG program has been altered with the employment of short characteristics and the application
of an accelerated lambda iteration (ALI) scheme.  Original development of the short characteristics (SC) methodology in the context of radiative transfer 
was done by Mihalas, Auer, \& Mihalas (1978\citep{mih78}).  Improvement of the SC approach in the explicit specification of the source function 
(at all grid points) was made by Olson \& Kunasz (1987)\citep{ols87} and Kunasz \& Auer (1988\citep{kun88}).  As a supplemental reference, the current 
version of MOOG draws upon the concise treatment of Auer \& Paletou (1994\citep{aue94}).  The implementation of the ALI technique within the framework of 
radiative transfer and stellar atmospheres was first done by Werner (1986\citep{wer86}) and subsequently refined 
by both \eg, Rybicki \& Hummer (1991\citep{ryb91}) and Hubeny (1992\citep{hub92}).  The main SC and ALI prescription followed by the MOOG code is that 
from Koesterke et al. (2002\citep{koe02} and references therein).  Since the text below is a general description and pertains to the specific coding in MOOG, 
the reader should consult the aforementioned references as they contain significantly more information.

The primary modification to MOOG is the creation and incorporation of four new subroutines.  The names and purposes of the new subroutines are as follows: 
\texttt{AngWeight.f}, which determines the Gaussian weights and integration points; \texttt{Sourcefunc\us scat\us cont.f}, which incorporates both a 
scattering and an absorption component to compute the source function and resultant flux for the continuum; \texttt{Sourcefunc\us scat\us line.f}, which incorporates 
both a scattering and an absorption component to compute the source function and resultant flux for the line; 
and, \texttt{Cdcalc\us JS.f}, which calculates the final line depth via the emergent continuum and line fluxes.  To accommodate these additions, 
several key subroutines were also revised.  Further details and the publicly-available MOOG code may be found at the website: \url{http://verdi.as.utexas.edu/moog.html}.  
Note that MOOG still retains the capacity to operate in {\it pure absorption} mode with the source function set simply to $S = \epsilon B$.

\section {Contributions to the Continuous Opacity}

To remind the reader, in the visible spectral range, the two principal sources of opacity in stellar atmospheres are the bound-free absorption from the
negative hydrogen ion (\hminus) and Rayleigh scattering from neutral atomic hydrogen.  The standard expression
(\eg\ Gray 1976\citep{gra76}) for the \hminus$_{BF}$ absorption coefficient is

\begin{equation}
\kappa \simeq 4.1458 \times 10^{-10} \alpha_{BF} P_{e} \Theta^{5/2} 10^{0.754\Theta}
\label{eqAA}
\end{equation}

\noindent where $\alpha_{BF}$ is the bound-free atomic absorption coefficient (which has frequency dependence), $P_{e}$ is the
electron pressure, and $\Theta = 5040/T$ (note that Eq.~\ref{eqAA} is per neutral hydrogen atom).

With regard to Rayleigh scattering, the scattering cross-section of radiation with angular frequency $\omega$ incident upon a neutral H atom
is given by the Kramers-Heisenberg formula in terms of atomic units as

\begin{equation}
\frac {\sigma(\omega)}{\sigma_{T}}= \left( \frac{\omega}{\omega_{1}} \right)^{4} \left| A_{0}
                                    + A_{2} \left( \frac{\omega}{\omega_{1}} \right)^{2}
                                    + A_{4} \left( \frac{\omega}{\omega_{1}} \right)^{4}+ \cdots \right|^2
\label{eqAB}
\end{equation}

\noindent where $\sigma_{T}$ is the Thompson scattering cross-section and $\omega_{1}$ is the angular frequency corresponding
to the Lyman limit.  Numerical calculations of the $A_{i}$ coefficients and the generation of an exact expression
for Eq.~\ref{eqAB}\ have been done by Dalgarno \& Williams (1962\citep{dal62}) and more recently by Lee \& Kim (2004\citep{lee04}).

The \hminus$_{BF}$ and Rayleigh scattering opacity contributions depend on temperature and metallicity (and to some extent, on the 
surface gravity).  Rayleigh scattering {\it also} has a $\lambda^{-4}$ dependence, and as a consequence, it greatly influences blue wavelength transitions.  
For the majority of stars (such as dwarfs and sub-giants), \hminus$_{BF}$ is the dominant opacity source in the visible spectral regime.  
However for low temperature, low metallicity giants, the Rayleigh scattering contribution becomes comparable to and even exceeds that from \hminus$_{BF}$ in the
ultraviolet and blue visible wavelength regions.  Therefore, to accurately determine the line intensity with the correct amount of flux and opacity contribution
for {\it all} stellar types and over a {\it wide} spectral range, isotropic, coherent scattering must be considered.

\section {Form of the Radiative Transfer Equation and Implementation of the ALI Scheme}

The source function is then written as: $S = (1-\epsilon)J + \epsilon B$ (where $\epsilon$ is the thermal coupling parameter, 
$J$ is the mean intensity, and $B$ is the Planck function).  To commence with the formal solution of the radiation transfer equation, a 
fundamental assumption is made in that the source function is specified completely in terms of optical depth.  After some mathematical manipulation, the
radiative transfer equation becomes 

\begin{equation}
\mu^{2} \frac{d^{2}j}{d\tau^{2}} = J - S
\label{eqC}
\end{equation}

\noindent where $\mu$ is the directional cosine and $\tau$ is the optical depth.  Eq.~\ref{eqC} is 
an integro-differential equation (and subject to boundary conditions).  To obtain the numerical solution
of Eq.~\ref{eqC}, a discretization in angle and optical depth is necessary.  As a consequence, the solution is simplified and a 
Gaussian quadrature summation is done instead of an integration.  Though it is eventually possible to 
evaluate Eq.~\ref{eqC} in a single step, the use of an iterative method to arrive at a solution is preferred as it is computationally faster than a straightforward 
approach.  For the MOOG program, the ALI technique is employed with the application of a full acceleration.  In the context of ALI scheme, the transfer equation
takes the form of $J = \Lambda[S]$, where $\Lambda$ represents the matrix operator.  Through the concept of preconditioning, ALI allows for the 
efficient, iterative solution of a (potentially) large system of linear equations.  The main steps of the iterative cycle are the evaluation of $J = \Lambda[S]$, 
the computation of the $\Delta S$ quantity, and the corresponding adjustment to the source function.     

\section{Short Characteristic Solution of Line Transfer}

From a general viewpoint, radiative transfer can be thought of as the propagation of photons along a ray on a two-dimensional
grid.  Note that the number of rays corresponds to the number of quadrature angles.  Determination of radiation along a ray is
done periodically at ray segments, or short characteristics.  Essentially, SC start at a grid point and proceed along the ray until a cell boundary is met.
At these cell boundaries, the intensity, opacity, and source function values are established.  Then with the knowledge of the cell 
boundary intensities, the intensity at other non-grid points can be calculated.  

Specifically with regard to MOOG, the intensity determination is a function of depths ($i$) and angles ($j$).  It is performed for both an inward ($i-1$) and 
an outward ($i+1$) direction.  Along the characteristic, the opacity quantity is assumed to be a linear function.  In effect, the intensity for the ray can
be expressed as

\begin{equation}
I = Ie^{-\Delta \tau(i)} + \int S(\tau)e^{-\tau}d\tau
\label{eqD}
\end{equation}

The optical depth step $\Delta\tau$ can be thought of as the path integral of the opacity along the characteristic (the entire, involved definition of the $\Delta\tau$ 
quantity is found in the \texttt{Sourcefunc\us scat\us *} subroutines).  Now, the evaluation of Eq.~\ref{eqD} requires the interpolation of the source function.  A 
linear interpolation is sufficient to satisfy the various boundary conditions.  Interpolation over [$S(i)$, $S(i \mp 1)$] then entails

\begin{equation}
\int S(\tau)e^{-\tau}d\tau = S(i)w_{1}(i) + S(i \mp 1)w_{2}(i)
\label{eqE}
\end{equation}

The weights are given by the relations

\begin{eqnarray}
w_{0}(i) = (e^{-\Delta\tau(i)}-1)/\Delta\tau(i)\\
w_{1}(i) = 1 + w_{0}(i)\\
w_{2}(i) = -e^{-\Delta\tau(i)}-w_{0}(i)
\label{eqF}
\end{eqnarray}

\noindent These weights are found by recursion.  The use of linear interpolation does not generate significant error (as normally would occur) 
due to the optically-thin nature of the boundary layer.  The expression for the mean intensity, $J$, subsequently becomes

\begin{equation}
J(i) = J(i) + 0.5w_{Gau}(j)I
\label{eqG}
\end{equation}

\noindent where the $w_{Gau}$ are the Gaussian quadrature weights (these are distinct from the weights of Eq.~\ref{eqE}).  The summation 
over all depth points and angles(/rays) is necessarily performed.  The SC formal solution of the transfer equation then proceeds in an iterative manner.  

Ongoing and future improvements to the MOOG code include: the incorporation of the Lee \& Kim (2004\citep{lee04}) formulation for
Rayleigh scattering (for atomic hydrogen), the employment of a further discretization with regard to frequency, and the 
implementation of spherically-symmetric geometry in the solution of radiative transfer.
 
\clearpage

\bibliography{JSobeck_M15_Bib}
\bibliographystyle{plain}


\clearpage

\clearpage

\clearpage

\clearpage

\clearpage

\clearpage

\clearpage

\clearpage

\clearpage

\clearpage

\clearpage

\clearpage


\clearpage
\begin{deluxetable}{lcccccccccccc}
\rotate
\tablenum{1}
\tablecolumns{13}
\tablewidth{0pt}
\tabletypesize{\scriptsize}

\tablecaption{Photometry and Model Atmosphere Parameters\label{m15models}}

\tablehead{
\colhead{}                          &
\colhead{}                          &                           
\colhead{}                          &
\colhead{Current-PHOTO\tablenotemark{a}} &
\colhead{}                          &
\colhead{Current-SPEC}              &                          
\colhead{}                          &
\colhead{}                          &
\colhead{}                          &
\colhead{Previous-SPEC\tablenotemark{b}} &
\colhead{}                          &
\colhead{}                          &                           
\colhead{}                          \\
\colhead{Star}                      &
\colhead{$V$\tablenotemark{c}}      &
\colhead{$V-K$\tablenotemark{d}}    &
\colhead{T$_{eff}$ [K]}              &
\colhead{$log(g)$}                  &
\colhead{T$_{eff}$ [K]}              &
\colhead{$log(g)$}                  &
\colhead{[M/H]}                     &
\colhead{v$_{turb}$ [km/s]}          &
\colhead{T$_{eff}$ [K]}              &
\colhead{$log(g)$}                  &
\colhead{[M/H]}                     &
\colhead{v$_{turb}$ [km/s]}         
}
\startdata
K341 & 12.81 & 3.115 & 4343 & 0.88 & 4375 & 0.30 & -2.25 & 2.00 &  4200 & 0.15 & -2.20 & 2.15 \\              
     &       &       &      &      &      &      &       &      &  4225 & 0.30 & -2.20 & 1.85 \\
K462 & 12.90 & 3.178 & 4298 & 0.89 & 4400 & 0.30 & -2.25 & 2.00 &  4225 & 0.15 & -2.20 & 2.10 \\
     &       &       &      &      &      &      &       &      &  4275 & 0.45 & -2.20 & 2.00 \\
K583 & 12.83 & 3.261 & 4241 & 0.82 & 4375 & 0.30 & -2.25 & 1.90 &  4275 & 0.15 & -2.20 & 2.20 \\              
     &       &       &      &      &      &      &       &      &  4275 & 0.30 & -2.20 & 1.90 \\
B009 & 15.20 & 2.091 & 5443 & 2.26 & 5300 & 1.65 & -2.50 & 2.60 &  5300 & 1.65 & -2.30 & 2.50 \\
B028 & 15.88 & 1.907 & 5718 & 2.65 & 5750 & 2.40 & -2.50 & 2.85 &  5750 & 2.40 & -2.50 & 2.75 \\
B224 & 15.69 & 1.775 & 5932 & 2.65 & 5600 & 2.10 & -2.50 & 2.60 &  5600 & 2.10 & -2.50 & 2.60 \\
B262 & 15.28 & 2.330 & 5122 & 2.16 & 4950 & 1.30 & -2.50 & 1.90 &  5000 & 1.50 & -2.30 & 1.70 \\
B412 & 15.81 & 1.888 & 5748 & 2.63 & 6200 & 2.70 & -2.50 & 3.30 &  6250 & 2.70 & -2.50 & 3.30 \\
B584 & 15.53 & 1.655 & 6142 & 2.66 & 6000 & 2.70 & -2.50 & 2.90 &  6000 & 2.70 & -2.30 & 2.70 \\
\enddata
\tablenotetext{a}{These values are derived with the IRFM effective temperature formulations (Eq.'s 8 and 9) of Alonso et al. 
                  (1999).}
\tablenotetext{b}{The previous values for the spectroscopic parameters for the RGB stars are taken from Sneden et al. (2000a, 1997) respectively
                and those for the RHB stars are obtained from Preston et al. (2006).}
\tablenotetext{c}{The RGB V magnitues are taken from Cudworth (2011) and those for the RHB targets are
                obtained from Buonanno et al. (1983).}
\tablenotetext{d}{The K magnitudes for all members of the M15 data set are taken from the 2MASS Database (Skrutskie et al. 2006).}
\end{deluxetable}

\clearpage
\begin{deluxetable}{lcccccccccc}
\rotate
\tablenum{2}
\tablecolumns{10}
\tablewidth{0pt}
\tabletypesize{\scriptsize}

\tablecaption{Element Abundances for Individual Transitions in M15 RGB Stars\label{m15rgbdata}}

\tablehead{
\colhead{Element/}                  &
\colhead{Species}                   &
\colhead{Measurement}               &
\colhead{Wavelength}                &
\colhead{$\chi$}                    &
\colhead{$log(gf)$}                 &
\colhead{$log\epsilon_{K341}$}       &
\colhead{$log\epsilon_{K462}$}       &
\colhead{$log\epsilon_{K583}$}       &
\colhead{$gf$}                      \\
\colhead{Ionization State}          &
\colhead{}                          &
\colhead{Type}                      &
\colhead{[\AA]}                     &
\colhead{[eV]}                      &
\colhead{}                          &
\colhead{}                          &
\colhead{}                          &
\colhead{}                          &
\colhead{Reference}                 \\
}
\startdata
C I (CH)	&	6.0	&	SYNTH	&	$\approx$ 4300.000	&	\nodata	&	\nodata	&	5.51	&	5.76	&	5.41	&	Plez \& Cohen 2005	\\
O I	&	8.0	&	SYNTH	&	6300.304	&	0.000	&	-9.776	&	7.03	&	6.88	&	6.75	&	Baluja \& Zeippen 1988	\\
O I	&	8.0	&	SYNTH	&	6363.776	&	0.020	&	-10.257	&	7.01	&	7.01	&	\nodata	&	Baluja \& Zeippen 1988	\\
Na I	&	11.0	&	EW	&	5889.950	&	0.000	&	0.108	&	3.45	&	3.61	&	4.70	&	Volz et al. 1996	\\
Na I	&	11.0	&	EW	&	5895.920	&	0.000	&	-0.194	&	\nodata	&	3.42	&	4.70	&	Volz et al. 1996	\\
Mg I	&	12.0	&	EW	&	4571.100	&	0.000	&	-5.623	&	5.39	&	5.42	&	5.33	&	Tachiev \& Fischer 2003	\\
Mg I	&	12.0	&	EW	&	4702.990	&	4.346	&	-0.440	&	5.65	&	5.61	&	5.43	&	Chang \& Tang 1990	\\
Mg I	&	12.0	&	EW	&	5528.400	&	4.346	&	-0.498	&	5.72	&	5.35	&	5.48	&	Chang \& Tang 1990	\\
Mg I	&	12.0	&	EW	&	5711.088	&	4.346	&	-1.724	&	\nodata	&	5.34	&	5.49	&	Chang \& Tang 1990	\\
Al I	&	13.0	&	EW	&	3944.010	&	0.000	&	-0.638	&	4.18	&	4.23	&	4.54	&	Mendoza et al. 1995	\\
Al I	&	13.0	&	EW	&	3961.520	&	0.010	&	-0.336	&	3.93	&	3.99	&	4.31	&	Mendoza et al. 1995	\\
Si I	&	14.0	&	EW	&	3905.523	&	1.910	&	-1.041	&	5.11	&	\nodata	&	5.42	&	O'Brian \& Lawler 1991	\\
Si I	&	14.0	&	EW	&	4102.936	&	1.910	&	-3.336	&	5.51	&	5.76	&	5.61	&	Fischer 2004	\\
Si I	&	14.0	&	EW	&	5684.484	&	4.954	&	-1.420	&	5.16	&	\nodata	&	\nodata	&	Nahar \& Padhan 1993	\\
Si I	&	14.0	&	EW	&	5690.425	&	4.930	&	-1.870	&	5.43	&	\nodata	&	5.40	&	Garz 1973	\\
Ca I	&	20.0	&	EW	&	4425.440	&	1.879	&	-0.358	&	3.91	&	4.02	&	3.87	&	Fuhr \& Wiese 1998	\\
Ca I	&	20.0	&	EW	&	5588.760	&	2.526	&	0.210	&	3.94	&	3.83	&	3.89	&	Kostlin 1964	\\
Ca I	&	20.0	&	EW	&	5857.460	&	2.933	&	0.230	&	4.07	&	3.99	&	4.06	&	Kostlin 1964	\\
Ca I	&	20.0	&	EW	&	6102.727	&	1.879	&	-0.790	&	\nodata	&	4.15	&	3.97	&	Fuhr \& Wiese 1998	\\
Ca I	&	20.0	&	EW	&	6122.230	&	1.886	&	-0.315	&	4.19	&	4.00	&	4.03	&	Fuhr \& Wiese 1998	\\
Ca I	&	20.0	&	EW	&	6162.180	&	1.900	&	-0.089	&	4.19	&	3.93	&	4.17	&	Fuhr \& Wiese 1998	\\
Ca I	&	20.0	&	EW	&	6439.070	&	2.526	&	0.470	&	\nodata	&	4.02	&	4.15	&	Kostlin 1964	\\
Ca I	&	20.0	&	EW	&	6717.690	&	2.709	&	-0.524	&	4.29	&	4.14	&	4.13	&	Fuhr \& Wiese 1998	\\

\enddata
\tablecomments{Table 2 is published in its entirety in the electronic edition of the
{\it Astronomical Journal}.  A portion is shown here for guidance with regard to its form and content.}

\end{deluxetable}

\clearpage
\begin{deluxetable}{lccccccccccccccccc}
\rotate
\tablenum{3}
\tablecolumns{13}
\tablewidth{0pt}
\tabletypesize{\scriptsize}

\tablecaption{Element Abundances for Individual Transitions in M15 RHB Stars\label{m15rhbdata}}

\tablehead{
\colhead{Element/}                  &
\colhead{Species}                   &
\colhead{Measurement}               &
\colhead{$\lambda$}                 & 
\colhead{$\chi$}                    &
\colhead{$log(gf)$}                 &
\colhead{$log\epsilon_{B009}$}       &
\colhead{$log\epsilon_{B028}$}       &
\colhead{$log\epsilon_{B224}$}       &
\colhead{$log\epsilon_{B262}$}       &
\colhead{$log\epsilon_{B412}$}       &
\colhead{$log\epsilon_{B584}$}       &
\colhead{$gf$}                      \\
\colhead{Ionization State}          &
\colhead{}                          &
\colhead{Type}                      &
\colhead{[\AA]}                     &
\colhead{[eV]}                      &
\colhead{}                          &
\colhead{}                          &
\colhead{}                          &
\colhead{}                          &
\colhead{}                          &
\colhead{}                          &
\colhead{}                          &
\colhead{Reference}                 \\
}
\startdata
Na I	&	11.0	&	EW	&	5889.950	&	0.000	&	0.112	&	5.09	&	4.05	&	4.67	&	4.63	&	3.81	&	3.86	&	Volz et al. 1996	\\
Na I	&	11.0	&	EW	&	5895.920	&	0.000	&	-0.191	&	5.07	&	3.88	&	4.54	&	4.58	&	3.79	&	3.75	&	Volz et al. 1996	\\
Mg I	&	12.0	&	EW	&	3829.360	&	2.710	&	-0.227	&	4.78	&	5.37	&	5.35	&	\nodata	&	\nodata	&	5.38	&	Tachiev \& Fischer 2003	\\
Mg I	&	12.0	&	EW	&	3832.310	&	2.710	&	-0.353	&	4.75	&	5.42	&	5.37	&	5.04	&	\nodata	&	5.16	&	Tachiev \& Fischer 2003	\\
Mg I	&	12.0	&	EW	&	4571.100	&	0.000	&	-5.623	&	4.90	&	5.31	&	5.26	&	5.19	&	5.34	&	5.57	&	Tachiev \& Fischer 2003	\\
Mg I	&	12.0	&	EW	&	5172.684	&	2.712	&	-0.393	&	4.73	&	5.61	&	5.44	&	5.34	&	5.57	&	5.52	&	Tachiev \& Fischer 2003	\\
Mg I	&	12.0	&	EW	&	5183.604	&	2.717	&	-0.167	&	4.91	&	5.62	&	\nodata	&	5.52	&	5.76	&	5.57	&	Tachiev \& Fischer 2003	\\
Mg I	&	12.0	&	EW	&	5528.420	&	4.346	&	-0.498	&	4.92	&	5.25	&	5.33	&	5.14	&	5.26	&	5.37	&	Chang \& Tang 1990	\\
Al I	&	13.0	&	EW	&	3944.010	&	0.000	&	-0.638	&	4.09	&	3.21	&	3.42	&	3.50	&	3.31	&	3.22	&	Mendoza et al. 1995	\\
Al I	&	13.0	&	EW	&	3961.530	&	0.014	&	-0.336	&	3.94	&	3.01	&	3.30	&	3.19	&	3.16	&	3.13	&	Mendoza et al. 1995	\\
Si I	&	14.0	&	EW	&	3905.523	&	1.910	&	-1.041	&	\nodata	&	5.11	&	5.30	&	5.29	&	4.98	&	4.87	&	O'Brian \& Lawler 1991	\\
Si I	&	14.0	&	EW	&	4102.936	&	1.910	&	-3.336	&	5.24	&	\nodata	&	\nodata	&	\nodata	&	\nodata	&	\nodata	&	Fischer 2004	\\
Ca I	&	20.0	&	EW	&	4226.740	&	0.000	&	0.244	&	4.05	&	4.29	&	4.36	&	3.70	&	4.29	&	4.22	&	Fuhr \& Wiese 1998	\\
Ca I	&	20.0	&	EW	&	5588.760	&	2.526	&	0.210	&	4.21	&	4.02	&	4.12	&	3.97	&	4.18	&	\nodata	&	Kostlin 1964	\\
Ca I	&	20.0	&	EW	&	5857.460	&	2.933	&	0.230	&	4.25	&	\nodata	&	4.14	&	4.24	&	\nodata	&	\nodata	&	Kostlin 1964	\\
Ca I	&	20.0	&	EW	&	6102.727	&	1.879	&	-0.790	&	4.19	&	\nodata	&	\nodata	&	3.97	&	4.27	&	\nodata	&	Fuhr \& Wiese 1998	\\
Ca I	&	20.0	&	EW	&	6122.230	&	1.886	&	-0.315	&	4.22	&	4.22	&	4.14	&	4.12	&	4.15	&	\nodata	&	Fuhr \& Wiese 1998	\\
Ca I	&	20.0	&	EW	&	6162.180	&	1.900	&	-0.089	&	4.29	&	4.20	&	4.08	&	4.20	&	4.20	&	4.13	&	Fuhr \& Wiese 1998	\\
Ca I	&	20.0	&	EW	&	6439.070	&	2.526	&	0.470	&	4.01	&	4.26	&	4.01	&	4.05	&	4.08	&	\nodata	&	Kostlin 1964	\\
Ca I	&	20.0	&	EW	&	6493.780	&	2.521	&	0.140	&	3.91	&	\nodata	&	3.97	&	\nodata	&	\nodata	&	\nodata	&	Kostlin 1964	\\
Ca I	&	20.0	&	EW	&	6499.650	&	2.523	&	-0.590	&	4.19	&	\nodata	&	\nodata	&	\nodata	&	\nodata	&	\nodata	&	Kostlin 1964	\\
Ca I	&	20.0	&	EW	&	6717.690	&	2.709	&	-0.524	&	\nodata	&	\nodata	&	\nodata	&	4.24	&	\nodata	&	\nodata	&	Fuhr \& Wiese 1998	\\

\enddata
\tablecomments{Table 3 is published in its entirety in the electronic edition of the
{\it Astronomical Journal}.  A portion is shown here for guidance with regard to its form and content.}

\end{deluxetable}

\clearpage
\begin{deluxetable}{lccccccccccccccccc}
\tablenum{4}
\tablecolumns{10}
\tablewidth{0pt}
\tabletypesize{\footnotesize}

\tablecaption{Average Relative Abundance Data for M15 Program Stars\label{m15means}}

\tablehead{
\colhead{}           &
\colhead{K341}       &
\colhead{K462}       &
\colhead{K583}       &
\colhead{B009}       &
\colhead{B028}       &
\colhead{B224}       &
\colhead{B262}       &
\colhead{B412}       &
\colhead{B584}       \\
}
\startdata
$<$[C$_{I}$/Fe$_{I}$]$>$	&	-0.34	&	-0.08	&	-0.40	&	-0.10	&	\nodata	&	\nodata	&	\nodata	&	\nodata	&	\nodata	\\
$\sigma$ (log$\epsilon$(C$_{I}$))	&	\nodata	&	\nodata	&	\nodata	&	\nodata	&	\nodata	&	\nodata	&	\nodata	&	\nodata	&	\nodata	\\
No. of Lines	&	1	&	1	&	1	&	1	&	0	&	0	&	0	&	0	&	0	\\
$<$[O$_{I}$/Fe$_{I}$]$>$	&	0.85	&	0.78	&	0.62	&	\nodata	&	\nodata	&	\nodata	&	\nodata	&	\nodata	&	\nodata	\\
$\sigma$ (log$\epsilon$(O$_{I}$))	&	0.01	&	0.09	&	\nodata	&	\nodata	&	\nodata	&	\nodata	&	\nodata	&	\nodata	&	\nodata	\\
No. of Lines	&	2	&	2	&	1	&	0	&	0	&	0	&	0	&	0	&	0	\\
$<$[Na$_{I}$/Fe$_{I}$]$>$	&	-0.18	&	-0.11	&	1.11	&	1.60	&	0.46	&	1.18	&	1.21	&	0.25	&	0.28	\\
$\sigma$ (log$\epsilon$(Na$_{I}$))	&	\nodata	&	0.13	&	0.00	&	0.01	&	0.12	&	0.09	&	0.04	&	0.01	&	0.08	\\
No. of Lines	&	1	&	2	&	2	&	2	&	2	&	2	&	2	&	2	&	2	\\
$<$[Mg$_{I}$/Fe$_{I}$]$>$	&	0.60	&	0.45	&	0.48	&	-0.01	&	0.56	&	0.56	&	0.44	&	0.57	&	0.54	\\
$\sigma$ (log$\epsilon$(Mg$_{I}$))	&	0.17	&	0.13	&	0.07	&	0.09	&	0.15	&	0.07	&	0.20	&	0.23	&	0.16	\\
No. of Lines	&	3	&	4	&	4	&	6	&	2	&	1	&	2	&	2	&	2	\\
$<$[Al$_{I}$/Fe$_{I}$]$>$	&	0.23	&	0.29	&	0.64	&	0.34	&	-0.60	&	-0.27	&	-0.26	&	-0.52	&	-0.52	\\
$\sigma$ (log$\epsilon$(Al$_{I}$))	&	0.18	&	0.17	&	0.16	&	0.11	&	0.14	&	0.08	&	0.22	&	0.11	&	0.06	\\
No. of Lines	&	2	&	2	&	2	&	2	&	2	&	2	&	2	&	2	&	2	\\
$<$[Si$_{I}$/Fe$_{I}$]$>$	&	0.33	&	0.80	&	0.55	&	0.42	&	0.26	&	0.53	&	0.55	&	0.09	&	0.00	\\
$\sigma$ (log$\epsilon$(Si$_{I}$))	&	0.20	&	0.08	&	0.01	&	\nodata	&	\nodata	&	\nodata	&	\nodata	&	\nodata	&	\nodata	\\
No. of Lines	&	4	&	2	&	2	&	1	&	1	&	1	&	1	&	1	&	1	\\
$<$[Ca$_{I}$/Fe$_{I}$]$>$	&	0.33	&	0.25	&	0.30	&	0.53	&	0.55	&	0.55	&	0.54	&	0.51	&	0.51	\\
$\sigma$ (log$\epsilon$(Ca$_{I}$))	&	0.15	&	0.10	&	0.12	&	0.13	&	0.11	&	0.13	&	0.18	&	0.08	&	0.06	\\
No. of Lines	&	6	&	8	&	8	&	9	&	5	&	7	&	9	&	6	&	2	\\
$<$[Sc$_{I}$/Fe$_{I}$]$>$	&	\nodata	&	\nodata	&	-0.76	&	\nodata	&	\nodata	&	\nodata	&	\nodata	&	\nodata	&	\nodata	\\
$\sigma$ (log$\epsilon$(Sc$_{I}$))	&	\nodata	&	\nodata	&	\nodata	&	\nodata	&	\nodata	&	\nodata	&	\nodata	&	\nodata	&	\nodata	\\
No. of Lines	&	0	&	0	&	1	&	0	&	0	&	0	&	0	&	0	&	0	\\
$<$[Sc$_{II}$/Fe$_{II}$]$>$	&	0.16	&	0.04	&	0.12	&	0.37	&	0.43	&	0.37	&	0.16	&	0.39	&	0.32	\\
$\sigma$ (log$\epsilon$(Sc$_{II}$))	&	0.10	&	0.12	&	0.10	&	0.16	&	0.06	&	0.08	&	0.09	&	0.20	&	\nodata	\\
No. of Lines	&	2	&	2	&	2	&	2	&	3	&	3	&	4	&	2	&	1	\\
$<$[Ti$_{I}$/Fe$_{I}$]$>$	&	0.11	&	0.19	&	0.07	&	0.58	&	0.60	&	0.54	&	0.32	&	0.85	&	0.77	\\
$\sigma$ (log$\epsilon$(Ti$_{I}$))	&	0.12	&	0.13	&	0.11	&	0.16	&	0.16	&	0.15	&	0.15	&	0.08	&	0.14	\\
No. of Lines	&	15	&	12	&	13	&	2	&	4	&	4	&	6	&	2	&	2	\\
$<$[Ti$_{II}$/Fe$_{II}$]$>$	&	0.47	&	0.44	&	0.47	&	0.38	&	0.48	&	0.40	&	0.31	&	0.52	&	0.50	\\
$\sigma$ (log$\epsilon$(Ti$_{II}$))	&	0.11	&	0.13	&	0.14	&	0.10	&	0.10	&	0.10	&	0.07	&	0.07	&	0.06	\\
No. of Lines	&	14	&	8	&	16	&	9	&	7	&	9	&	11	&	8	&	7	\\
$<$[V$_{I}$/Fe$_{I}$]$>$	&	-0.26	&	-0.11	&	-0.18	&	\nodata	&	\nodata	&	\nodata	&	\nodata	&	\nodata	&	\nodata	\\
$\sigma$ (log$\epsilon$(V$_{I}$))	&	0.07	&	0.19	&	0.12	&	\nodata	&	\nodata	&	\nodata	&	\nodata	&	\nodata	&	\nodata	\\
No. of Lines	&	2	&	3	&	3	&	0	&	0	&	0	&	0	&	0	&	0	\\
$<$[V$_{II}$/Fe$_{II}$]$>$	&	0.27	&	0.43	&	0.11	&	0.23	&	0.41	&	0.31	&	0.04	&	\nodata	&	0.44	\\
$\sigma$ (log$\epsilon$(V$_{II}$))	&	0.01	&	0.21	&	0.21	&	\nodata	&	0.05	&	\nodata	&	0.04	&	\nodata	&	0.03	\\
No. of Lines	&	2	&	2	&	2	&	1	&	2	&	1	&	2	&	0	&	2	\\
$<$[Cr$_{I}$/Fe$_{I}$]$>$	&	-0.36	&	-0.22	&	-0.38	&	-0.21	&	-0.24	&	-0.21	&	-0.22	&	-0.07	&	-0.19	\\
$\sigma$ (log$\epsilon$(Cr$_{I}$))	&	0.05	&	0.12	&	0.07	&	0.15	&	0.05	&	0.12	&	0.21	&	0.10	&	0.09	\\
No. of Lines	&	5	&	6	&	5	&	4	&	3	&	3	&	8	&	3	&	2	\\
$<$[Cr$_{II}$/Fe$_{II}$]$>$	&	-0.06	&	-0.02	&	-0.03	&	0.33	&	0.37	&	0.24	&	0.10	&	0.43	&	0.38	\\
$\sigma$ (log$\epsilon$(Cr$_{II}$))	&	0.14	&	0.18	&	0.06	&	0.01	&	0.21	&	\nodata	&	0.10	&	0.06	&	0.07	\\
No. of Lines	&	4	&	3	&	3	&	2	&	3	&	1	&	3	&	2	&	2	\\
$<$[Mn$_{I}$/Fe$_{I}$]$>$	&	-0.34	&	-0.45	&	-0.42	&	-0.70	&	-0.74	&	-0.80	&	-0.78	&	\nodata	&	\nodata	\\
$\sigma$ (log$\epsilon$(Mn$_{I}$))	&	0.02	&	0.04	&	\nodata	&	\nodata	&	\nodata	&	\nodata	&	\nodata	&	\nodata	&	\nodata	\\
No. of Lines	&	2	&	2	&	1	&	1	&	1	&	1	&	1	&	0	&	0	\\
$<$[Fe$_{I}$/H]$>$	&	-2.53	&	-2.54	&	-2.57	&	-2.69	&	-2.65	&	-2.74	&	-2.77	&	-2.62	&	-2.63	\\
$\sigma$ (log$\epsilon$(Fe$_{I}$))	&	0.14	&	0.12	&	0.14	&	0.12	&	0.12	&	0.11	&	0.14	&	0.12	&	0.15	\\
No. of Lines	&	59	&	57	&	63	&	98	&	72	&	72	&	100	&	45	&	60	\\
$<$[Fe$_{II}$/H]$>$	&	-2.51	&	-2.51	&	-2.57	&	-2.65	&	-2.60	&	-2.66	&	-2.73	&	-2.59	&	-2.59	\\
$\sigma$ (log$\epsilon$(Fe$_{II}$))	&	0.13	&	0.11	&	0.10	&	0.09	&	0.07	&	0.13	&	0.13	&	0.07	&	0.13	\\
No. of Lines	&	8	&	9	&	10	&	12	&	8	&	9	&	11	&	6	&	9	\\
$<$[Co$_{I}$/Fe$_{I}$]$>$	&	0.07	&	0.25	&	0.35	&	0.45	&	0.18	&	0.18	&	0.04	&	\nodata	&	0.22	\\
$\sigma$ (log$\epsilon$(Co$_{I}$))	&	0.21	&	0.18	&	0.20	&	0.32	&	\nodata	&	\nodata	&	\nodata	&	\nodata	&	\nodata	\\
No. of Lines	&	2	&	2	&	2	&	2	&	1	&	1	&	1	&	0	&	1	\\
$<$[Ni$_{I}$/Fe$_{I}$]$>$	&	\nodata	&	\nodata	&	\nodata	&	-0.23	&	\nodata	&	\nodata	&	\nodata	&	\nodata	&	\nodata	\\
$\sigma$ (log$\epsilon$(Ni$_{I}$))	&	\nodata	&	\nodata	&	\nodata	&	0.11	&	\nodata	&	\nodata	&	\nodata	&	\nodata	&	\nodata	\\
No. of Lines	&	0	&	0	&	0	&	2	&	0	&	0	&	0	&	0	&	0	\\
$<$[Cu$_{I}$/Fe$_{I}$]$>$	&	-0.92	&	-0.88	&	-0.92	&	\nodata	&	\nodata	&	\nodata	&	\nodata	&	\nodata	&	\nodata	\\
$\sigma$ (log$\epsilon$(Cu$_{I}$))	&	\nodata	&	\nodata	&	\nodata	&	\nodata	&	\nodata	&	\nodata	&	\nodata	&	\nodata	&	\nodata	\\
No. of Lines	&	1	&	1	&	1	&	0	&	0	&	0	&	0	&	0	&	0	\\
$<$[Zn$_{I}$/Fe$_{I}$]$>$	&	0.02	&	-0.01	&	0.01	&	\nodata	&	\nodata	&	0.49	&	0.27	&	\nodata	&	\nodata	\\
$\sigma$ (log$\epsilon$(Zn$_{I}$))	&	0.07	&	0.07	&	\nodata	&	\nodata	&	\nodata	&	\nodata	&	0.06	&	\nodata	&	\nodata	\\
No. of Lines	&	2	&	2	&	2	&	0	&	0	&	1	&	2	&	0	&	0	\\
$<$[Sr$_{II}$/Fe$_{II}$]$>$	&	-0.01	&	0.28	&	-0.18	&	0.16	&	0.35	&	0.43	&	0.06	&	0.38	&	0.46	\\
$\sigma$ (log$\epsilon$(Sr$_{II}$))	&	0.00	&	0.04	&	0.11	&	0.04	&	0.11	&	0.14	&	0.11	&	0.14	&	0.11	\\
No. of Lines	&	2	&	2	&	2	&	2	&	2	&	2	&	2	&	2	&	2	\\
$<$[Y$_{II}$/Fe$_{II}$]$>$	&	-0.13	&	-0.04	&	-0.23	&	0.19	&	0.02	&	0.16	&	-0.04	&	0.18	&	0.08	\\
$\sigma$ (log$\epsilon$(Y$_{II}$))	&	0.04	&	0.04	&	0.04	&	0.06	&	0.11	&	0.09	&	0.09	&	0.04	&	0.03	\\
No. of Lines	&	7	&	7	&	6	&	6	&	2	&	4	&	10	&	2	&	3	\\
$<$[Zr$_{II}$/Fe$_{II}$]$>$	&	0.21	&	0.27	&	0.14	&	0.53	&	0.35	&	0.38	&	0.35	&	\nodata	&	0.71	\\
$\sigma$ (log$\epsilon$(Zr$_{II}$))	&	0.06	&	0.12	&	0.17	&	0.16	&	\nodata	&	0.14	&	0.02	&	\nodata	&	0.00	\\
No. of Lines	&	4	&	4	&	4	&	4	&	1	&	2	&	5	&	0	&	2	\\
$<$[Ba$_{II}$/Fe$_{II}$]$>$	&	0.20	&	0.23	&	0.01	&	0.40	&	-0.14	&	0.21	&	0.43	&	0.14	&	-0.03	\\
$\sigma$ (log$\epsilon$(Ba$_{II}$))	&	0.03	&	0.10	&	0.07	&	0.08	&	0.03	&	0.05	&	0.10	&	0.07	&	0.06	\\
No. of Lines	&	4	&	4	&	4	&	5	&	3	&	5	&	6	&	3	&	3	\\
$<$[La$_{II}$/Fe$_{II}$]$>$	&	0.16	&	0.42	&	-0.04	&	0.60	&	-0.01	&	0.58	&	0.38	&		&	0.60	\\
$\sigma$ (log$\epsilon$(La$_{II}$))	&	0.08	&	0.09	&	0.09	&	0.07	&	\nodata	&	0.10	&	0.08	&	\nodata	&	0.14	\\
No. of Lines	&	9	&	9	&	9	&	8	&	1	&	6	&	11	&	1	&	2	\\
$<$[Ce$_{II}$/Fe$_{II}$]$>$	&	0.20	&	0.33	&	-0.17	&	0.51	&	\nodata	&	\nodata	&	0.37	&	\nodata	&	\nodata	\\
$\sigma$ (log$\epsilon$(Ce$_{II}$))	&	0.15	&	0.14	&	0.13	&	0.13	&	\nodata	&	\nodata	&	0.14	&	\nodata	&	\nodata	\\
No. of Lines	&	25	&	22	&	11	&	10	&	0	&	0	&	13	&	0	&	0	\\
$<$[Pr$_{II}$/Fe$_{II}$]$>$	&	0.42	&	0.63	&	0.28	&	0.73	&	\nodata	&	0.81	&	0.69	&	\nodata	&	\nodata	\\
$\sigma$ (log$\epsilon$(Pr$_{II}$))	&	0.04	&	0.03	&	0.12	&	0.00	&	\nodata	&	\nodata	&	0.11	&	\nodata	&	\nodata	\\
No. of Lines	&	5	&	5	&	5	&	2	&	0	&	1	&	2	&	0	&	0	\\
$<$[Nd$_{II}$/Fe$_{II}$]$>$	&	0.47	&	0.68	&	0.23	&	0.58	&	\nodata	&	\nodata	&	\nodata	&	\nodata	&	\nodata	\\
$\sigma$ (log$\epsilon$(Nd$_{II}$))	&	0.10	&	0.11	&	0.14	&	0.10	&	\nodata	&	\nodata	&	\nodata	&	\nodata	&	\nodata	\\
No. of Lines	&	27	&	25	&	22	&	2	&	0	&	0	&	0	&	0	&	0	\\
$<$[Sm$_{II}$/Fe$_{II}$]$>$	&	0.50	&	0.77	&	0.29	&	0.61	&	\nodata	&	\nodata	&	\nodata	&	\nodata	&	\nodata	\\
$\sigma$ (log$\epsilon$(Sm$_{II}$))	&	0.08	&	0.08	&	0.10	&	0.13	&	\nodata	&	\nodata	&	\nodata	&	\nodata	&	\nodata	\\
No. of Lines	&	20	&	20	&	15	&	3	&	0	&	0	&	0	&	0	&	0	\\
$<$[Eu$_{II}$/Fe$_{II}$]$>$	&	0.50	&	0.83	&	0.26	&	0.90	&	0.72	&	0.85	&	0.69	&	1.18	&	0.96	\\
$\sigma$ (log$\epsilon$(Eu$_{II}$))	&	0.12	&	0.11	&	0.11	&	0.03	&	0.04	&	0.09	&	0.07	&	0.07	&	0.18	\\
No. of Lines	&	7	&	6	&	7	&	5	&	2	&	5	&	5	&	2	&	2	\\
$<$[Gd$_{II}$/Fe$_{II}$]$>$	&	0.55	&	0.77	&	0.25	&	0.69	&	\nodata	&	\nodata	&	0.67	&	\nodata	&	\nodata	\\
$\sigma$ (log$\epsilon$(Gd$_{II}$))	&	0.08	&	0.07	&	0.03	&	\nodata	&	\nodata	&	\nodata	&	\nodata	&	\nodata	&	\nodata	\\
No. of Lines	&	8	&	9	&	3	&	1	&	0	&	0	&	1	&	0	&	0	\\
$<$[Tb$_{II}$/Fe$_{II}$]$>$	&	0.56	&	0.67	&	0.42	&	\nodata	&	\nodata	&	\nodata	&	0.96	&	\nodata	&	\nodata	\\
$\sigma$ (log$\epsilon$(Tb$_{II}$))	&	\nodata	&	\nodata	&	\nodata	&	\nodata	&	\nodata	&	\nodata	&	\nodata	&	\nodata	&	\nodata	\\
No. of Lines	&	1	&	1	&	1	&	0	&	0	&	0	&	1	&	0	&	0	\\
$<$[Dy$_{II}$/Fe$_{II}$]$>$	&	0.38	&	0.58	&	0.36	&	1.13	&	\nodata	&	\nodata	&	0.75	&	\nodata	&	\nodata	\\
$\sigma$ (log$\epsilon$(Dy$_{II}$))	&	0.15	&	0.03	&	0.07	&	0.08	&	\nodata	&	\nodata	&	0.11	&	\nodata	&	\nodata	\\
No. of Lines	&	4	&	2	&	4	&	3	&	0	&	0	&	3	&	0	&	0	\\
$<$[Ho$_{II}$/Fe$_{II}$]$>$	&	0.45	&	0.67	&	0.26	&	1.09	&	\nodata	&	1.35	&	0.81	&	\nodata	&	\nodata	\\
$\sigma$ (log$\epsilon$(Ho$_{II}$))	&	0.05	&	0.06	&	0.10	&	0.10	&	\nodata	&	\nodata	&	0.05	&	\nodata	&	\nodata	\\
No. of Lines	&	3	&	4	&	3	&	3	&	0	&	1	&	4	&	0	&	0	\\
$<$[Er$_{II}$/Fe$_{II}$]$>$	&	0.48	&	0.67	&	0.18	&	0.88	&	\nodata	&	0.73	&	0.58	&	\nodata	&	\nodata	\\
$\sigma$ (log$\epsilon$(Er$_{II}$))	&	0.15	&	0.15	&	0.06	&	0.03	&	\nodata	&	0.14	&	0.05	&	\nodata	&	\nodata	\\
No. of Lines	&	4	&	4	&	3	&	4	&	0	&	2	&	3	&	0	&	0	\\
$<$[Tm$_{II}$/Fe$_{II}$]$>$	&	0.40	&	0.66	&	0.26	&	\nodata	&	\nodata	&	\nodata	&	0.85	&	\nodata	&	\nodata	\\
$\sigma$ (log$\epsilon$(Tm$_{II}$))	&	0.09	&	0.20	&	0.07	&	\nodata	&	\nodata	&	\nodata	&	\nodata	&	\nodata	&	\nodata	\\
No. of Lines	&	3	&	3	&	2	&	0	&	0	&	0	&	1	&	0	&	0	\\
$<$[Yb$_{II}$/Fe$_{II}$]$>$	&	0.23	&	0.64	&	0.09	&	0.93	&	1.29	&	0.71	&	\nodata	&	0.89	&	0.71	\\
$\sigma$ (log$\epsilon$(Yb$_{II}$))	&	\nodata	&	\nodata	&	\nodata	&	\nodata	&	\nodata	&	\nodata	&	\nodata	&	\nodata	&	\nodata	\\
No. of Lines	&	1	&	1	&	1	&	1	&	1	&	1	&	0	&	1	&	1	\\
$<$[Hf$_{II}$/Fe$_{II}$]$>$	&	0.21	&	0.52	&	-0.08	&	\nodata	&	\nodata	&	\nodata	&	\nodata	&	\nodata	&	\nodata	\\
$\sigma$ (log$\epsilon$(Hf$_{II}$))	&	\nodata	&	\nodata	&	\nodata	&	\nodata	&	\nodata	&	\nodata	&	\nodata	&	\nodata	&	\nodata	\\
No. of Lines	&	1	&	1	&	1	&	0	&	0	&	0	&	0	&	0	&	0	\\
$<$[Os$_{I}$/Fe$_{I}$]$>$	&	\nodata	&	\nodata	&	\nodata	&	\nodata	&	\nodata	&	\nodata	&	2.00	&	\nodata	&	\nodata	\\
$\sigma$ (log$\epsilon$(Os$_{I}$))	&	\nodata	&	\nodata	&	\nodata	&	\nodata	&	\nodata	&	\nodata	&	0.35	&	\nodata	&	\nodata	\\
No. of Lines	&	0	&	0	&	0	&	0	&	0	&	0	&	2	&	0	&	0	\\
$<$[Ir$_{I}$/Fe$_{I}$]$>$	&	\nodata	&	\nodata	&	\nodata	&	\nodata	&	\nodata	&	\nodata	&	1.35	&	\nodata	&	\nodata	\\
$\sigma$ (log$\epsilon$(Ir$_{I}$))	&	\nodata	&	\nodata	&	\nodata	&	\nodata	&	\nodata	&	\nodata	&	\nodata	&	\nodata	&	\nodata	\\
No. of Lines	&	0	&	0	&	0	&	0	&	0	&	0	&	1	&	0	&	0	\\
$<$[Pb$_{I}$/Fe$_{I}$]$>$	&	$\lesssim$ 0.44	&	$\lesssim$ 0.35	&	$\lesssim$ 0.38	&	\nodata	&	\nodata	&	\nodata	&	\nodata	&	\nodata	&	\nodata	\\
$\sigma$ (log$\epsilon$(Pb$_{I}$))	&	\nodata	&	\nodata	&	\nodata	&	\nodata	&	\nodata	&	\nodata	&	\nodata	&	\nodata	&	\nodata	\\
No. of Lines	&	1	&	1	&	1	&	0	&	0	&	0	&	0	&	0	&	0	\\
$<$[Th$_{II}$/Fe$_{II}$]$>$	&	0.46	&	0.67	&	0.32	&	\nodata	&	\nodata	&	\nodata	&	\nodata	&	\nodata	&	\nodata	\\
$\sigma$ (log$\epsilon$(Th$_{II}$))	&	\nodata	&	\nodata	&	\nodata	&	\nodata	&	\nodata	&	\nodata	&	\nodata	&	\nodata	&	\nodata	\\
No. of Lines	&	1	&	1	&	1	&	0	&	0	&	0	&	0	&	0	&	0	\\

\enddata
\end{deluxetable}

\clearpage
\begin{deluxetable}{lccccc}
\tablenum{5}
\tablecolumns{5}
\tablewidth{0pt}
\tabletypesize{\scriptsize}

\tablecaption{Solar Photospheric Abundances of the Elements Analyzed in the Current Study\label{solarabund}}
\tablehead{
\colhead{Element}                   &
\colhead{Z}                         &
\colhead{log($\epsilon_{El}$)}       &
\colhead{Reference}                 &
\colhead{Methodology}               \\
}
\startdata
H	&	1	&	12.00	&	N/A	&	\nodata	\\
C	&	6	&	8.43 $\pm$ 0.05	&	Asplund et al. 2009     &	3-D; LTE \\
O	&	8	&	8.71 $\pm$ 0.05	&	Scott et al. 2009	&	3-D; LTE \\
Na	&	11	&	6.17 $\pm$ 0.04	&	Asplund et al. 2005	&	3-D; LTE \\
Mg	&	12	&	7.53 $\pm$ 0.09	&	Asplund et al. 2005	&	3-D; LTE \\
Al	&	13	&	6.37 $\pm$ 0.06	&	Asplund et al. 2005	&	3-D; LTE \\
Si	&	14	&	7.51 $\pm$ 0.04	&	Asplund et al. 2005	&	3-D; LTE \\
Ca	&	20	&	6.31 $\pm$ 0.04	&	Asplund et al. 2005	&	3-D; LTE \\
Sc      &	21	&	3.15 $\pm$ 0.04 &	Asplund et al. 2009; Grevesse et al. 2010 & 3-D; LTE \\
Ti	&	22	&	4.95 $\pm$ 0.05	&	Asplund et al. 2009; Grevesse et al. 2010 & 3-D; LTE \\
V       &	23	&	3.93 $\pm$ 0.08	&	Asplund et al. 2009; Grevesse et al. 2010 & 3-D; LTE \\
Cr	&	24	&	5.64 $\pm$ 0.02	&	Sobeck et al. 2007	&	1-D; LTE \\
Mn      &	25	&	5.43 $\pm$ 0.04	&	Asplund et al. 2009; Grevesse et al. 2010 & 3-D; LTE \\
Fe	&	26	&	7.52 $\pm$ 0.08	&	Sneden et al. 1991	&	1-D; LTE \\
Co	&	27	&	4.99 $\pm$ 0.08 &	Asplund et al. 2009; Grevesse et al. 2010 & 3-D; LTE \\
Ni	&	28	&	6.17 $\pm$ 0.02	&	Scott et al. 2009	&	3-D; LTE \\
Cu	&	29	&	4.19 $\pm$ 0.04	&	Asplund et al. 2009; Grevesse et al. 2010 & 3-D; LTE \\
Zn	&	30	&	4.60 $\pm$ 0.03	&	Biemont \& Godefroid 1980	&	1-D; LTE \\
Sr	&	38	&	2.92 $\pm$ 0.05	&	Barklem \& O'Mara 2000	&	1-D; LTE	\\
Y	&	39	&	2.21 $\pm$ 0.05	&	Asplund et al. 2009; Grevesse et al. 2010 & 3-D; LTE \\
Zr	&	40	&	2.58 $\pm$ 0.04	&	Asplund et al. 2009; Grevesse et al. 2010 & 3-D; LTE \\
Ba	&	56	&	2.17 $\pm$ 0.09	&	Asplund et al. 2009; Grevesse et al. 2010 & 3-D; LTE \\
La	&	57	&	1.13 $\pm$ 0.03	&	Lawler et al. 2001	&	1-D; LTE	\\
Ce	&	58	&	1.61 $\pm$ 0.01	&	Lawler et al.2009	&	1-D; LTE	\\
Pr	&	59	&	0.76 $\pm$ 0.02	&	Sneden et al. 2009	&	1-D; LTE	\\
Nd	&	60	&	1.45 $\pm$ 0.01	&	Den Hartog et al. 2003	&	1-D; LTE	\\
Sm	&	62	&	1.00 $\pm$ 0.01	&	Lawler et al. 2006	&	1-D; LTE	\\
Eu	&	63	&	0.52 $\pm$ 0.01	&	Lawler et al. 2001	&	1-D; LTE	\\
Gd	&	64	&	1.11 $\pm$ 0.01	&	Den Hartog et al. 2006	&	1-D; LTE	\\
Tb	&	65	&	0.28 $\pm$ 0.3	&	Lawler et al. 2001	&	1-D; LTE	\\
Dy	&	66	&	1.13 $\pm$ 0.02	&	Sneden et al. 2009	&	1-D; LTE	\\
Ho	&	67	&	0.51 $\pm$ 0.10	&	Lawler et al. 2004	&	1-D; LTE	\\
Er	&	68	&	0.96 $\pm$ 0.03	&	Lawler et al. 2008	&	1-D; LTE	\\
Tm	&	69	&	0.14 $\pm$ 0.02	&	Sneden et al. 2009	&	1-D; LTE	\\
Yb	&	70	&	0.86 $\pm$ 0.10	&	Sneden et al. 2009	&	1-D; LTE	\\
Hf	&	72	&	0.88 $\pm$ 0.08	&	Lawler et al. 2007	&	1-D; LTE	\\
Os	&	76	&	1.25 $\pm$ 0.11	&	Quinet et al. 2006	&	1-D; LTE	\\
Ir	&	77	&	1.38 $\pm$ 0.05	&	Youssef \& Khalil 1988	&	1-D; LTE	\\
Pb      &       82      &       1.75 $\pm$ 0.10 &       Asplund et al. 2009; Grevesse et al. 2010 & 3-D; LTE \\
Th      &       90      &       0.02 $\pm$ 0.10 &       Asplund et al. 2009; Grevesse et al. 2010 & 3-D; LTE \\
\enddata
\end{deluxetable}

\end{document}